\documentclass[aip,amsmath,amssymb,floatfix,citeautoscript,reprint]{revtex4-1}
\usepackage{cancel}
\usepackage{amsmath}
\usepackage{graphicx}
\usepackage{bm}
\usepackage{physics}
\usepackage[version=3]{mhchem}
\usepackage{txfonts}

\usepackage{color}
\usepackage[usenames,dvipsnames]{xcolor}
\definecolor{myblue}{rgb}{0,0,1}
\usepackage[breaklinks=true,colorlinks=true,linkcolor=myblue,urlcolor=myblue,citecolor=myblue]{hyperref}

\begin{document}
\title{Beyond Walkers in Stochastic Quantum Chemistry: Reducing Error using Fast Randomized Iteration}

\author{Samuel M. Greene}
\affiliation{Department of Chemistry and James Franck Institute, University of Chicago, Chicago, Illinois 60637, United States}
\author{Robert J. Webber}
\author{Jonathan Weare}
\email{weare@cims.nyu.edu}
\affiliation{Courant Institute of Mathematical Sciences, New York University, New York, New York 10012, United States}
\author{Timothy C. Berkelbach}
\email{tim.berkelbach@gmail.com}
\affiliation{Department of Chemistry, Columbia University, New York, New York 10027, United States}
\affiliation{Center for Computational Quantum Physics, Flatiron Institute, New York, New York 10010, United States}

\begin{abstract}
We introduce a family of methods for the full configuration interaction problem
in quantum chemistry, based on the fast randomized iteration (FRI) framework
[L.-H. Lim and J. Weare, SIAM Rev.~\textbf{59}, 547 (2017)]. These methods,
which we term ``FCI-FRI,'' stochastically impose sparsity during iterations of the
power method and can be viewed as a generalization of full configuration
interaction quantum Monte Carlo (FCIQMC) without walkers.  In addition to the
multinomial scheme commonly used to sample excitations in FCIQMC, we present a
systematic scheme where excitations are not sampled independently.  Performing
ground-state calculations on five small molecules at fixed cost, we find that
the systematic FCI-FRI scheme is 11 to 45 times more statistically
efficient than the multinomial FCI-FRI scheme, which is in turn 1.4 to 178 times more statistically efficient than the original FCIQMC algorithm.
\end{abstract}

\maketitle

\section{Introduction}
Deterministic approaches to treating strong correlation in interacting quantum systems are often rendered intractable by the exponential scaling of the size of the Hilbert space with the number of particles.\cite{Troyer2005} 
In contrast, quantum Monte Carlo (QMC) methods~\cite{Barker1979, Hammond1994, CalandraBuonaura1998, Maksym2005, Needs2010, Booth2014, Austin2012, Shepherd2014} can be computationally more efficient because they employ a sparse
representation of the wave function in this space, obtained via stochastic sampling. Methods that utilize a continuous
basis of configurations in real space have long existed, e.g. diffusion Monte
Carlo~\cite{Umrigar1993, Senatore1994, Kosztin1996, Foulkes2001,
Manten2001, Hairer2014}. The application of these methods to fermionic systems requires nodal constraints due to the antisymmetry of the wave function. This has motivated the development of discrete-space methods, e.g.~full configuration interaction QMC (FCIQMC) and
auxiliary-field QMC~\cite{Booth2009, Li2015, Alavi2016, Motta2018}, in which the antisymmetry is provided by a Slater determinant basis, thereby obviating the need to impose nodal
constraints on the wave function.\cite{Booth2009, Austin2012, Spencer2012,
Umrigar2015} A disadvantage of discrete-basis methods is that the basis is not complete, but this can be
addressed using standard extrapolation techniques.\cite{Halkier1998,
Halkier1999}

Recently, Lim and Weare \cite{Lim2017} introduced the fast randomized iteration
(FRI) framework, a class of methods that use techniques similar to those used in
discrete-basis QMC methods to solve large, generic linear algebra problems. Sparsity is
imposed stochastically in matrices and vectors, which reduces the computational
cost and storage requirements of these methods and facilitates their application
to problems significantly larger than those treatable by conventional linear
algebra approaches. Many existing QMC algorithms, including the FCIQMC method, can be understood as specific methods within the FRI framework. The central purpose of this work is to describe, in a more general context, the application of FRI methods to calculations on interacting fermionic systems in a discrete basis. Importantly, we leverage this generality to develop alternative methods within this framework and investigate their statistical error and convergence properties through numerical tests on small molecular systems.

The FRI framework can be applied in a variety of ways to calculate ground- and excited-state observables of electronic systems. This study discusses only the application of FRI to calculate the ground-state energy of the full configuration interaction (FCI) Hamiltonian matrix in a Slater determinant basis. Such applications of the FRI framework will be referred to in this manuscript as FCI-FRI. In these methods, calculation of the ground-state energy is achieved via stochastic implementations of the power method, in which an
initial trial vector is evolved towards the ground state eigenvector by
repeatedly applying the Hamiltonian, scaled and shifted such that the ground state is
dominant. The power method can be viewed as a discretization of the
imaginary-time propagation used in many QMC methods. In order to reduce computational cost, the Hamiltonian matrix and solution vector are compressed
stochastically, meaning that randomly selected subsets of their elements are
zeroed in each iteration. Calculating the energy after each iteration and
averaging yields an estimate of the ground-state energy. This estimate can be
systematically improved by executing more iterations and by retaining more
nonzero elements in each compression.
Unlike the original FCIQMC method, some FRI methods become identical to the deterministic power method as the number of randomly selected elements increases to the size of the basis.

The various approaches to matrix and vector compression within the FRI framework differ in terms of their computational cost and statistical efficiency. In this study, we combine these approaches in two new FCI-FRI methods and compare them to the original FCIQMC method.\cite{Booth2009} In the first method, multinomial matrix compression, which is used in FCIQMC, is combined with systematic vector compression. Multinomial and systematic sampling are reviewed in Section~\ref{sec:samplScheme}. In the original presentation of FRI~\cite{Lim2017}, systematic vector compression was shown to yield the least statistical error out of all other schemes considered. In contrast, vector compression is achieved by integerizing elements in FCIQMC. Comparing the original FCIQMC method to the ``multinomial FCI-FRI'' method, which uses the same matrix compression scheme, illustrates the gains in efficiency that an improved vector compression scheme can enable. In the second method, ``systematic FCI-FRI,'' we seek to further improve the efficiency by also compressing the matrix systematically instead of multinomially.  We introduce a new hierarchical scheme to reduce the computational cost of performing this compression. In numerical tests on five small molecules, we find that systematic FCI-FRI yields consistently greater statistical efficiency (defined below) than multinomial FCI-FRI by at least an order of magnitude, and multinomial FCI-FRI is also more statistically efficient than FCIQMC in its original form.

An additional purpose of this work is to better understand how the features of each of these methods influence their errors and computational cost. To this end, we also compare two methods applied recently to FCI problems~\cite{Lu2017} in which the matrix is not compressed. Although expensive, such approaches are feasible because of the sparse structure of the Hamiltonian.  In the first of these methods, the vector is compressed using the stochastic systematic scheme, whereas in the second, it is compressed using a deterministic thresholding scheme. Both methods have similar cost and are tractable for problems beyond the reach of deterministic FCI. However, the stochastic method achieves significantly less error,  highlighting  the advantages of stochastic methods over their deterministic counterparts.

A number of recent extensions to the original FCIQMC algorithm have been found to enable improvements in performance by orders of magnitude. For example, in semi-stochastic FCIQMC\cite{Petruzielo2012, Blunt2015b}, a fixed subspace within the Slater determinant basis is treated deterministically, greatly reducing the statistical error in that portion of the solution vector. A related extension involves preserving some elements exactly if their magnitude exceeds a user-specified threshold~\cite{Overy2014}. In the initiator approximation\cite{Cleland2010, Booth2011, Cleland2011}, elements in the solution vector are zeroed in each iteration according to deterministic compression rules to better constrain the sign structure of the solution vector, which introduces a small bias. The FCI-FRI methods discussed here also include some deterministic features, although these differ in key aspects from those in the FCIQMC extensions. In FCI-FRI, the vector and matrix elements elements to be preserved exactly are chosen dynamically in each iteration on the basis of their relative magnitudes. The criteria for selecting these elements do not rely on user-specified parameters and instead were chosen to minimize compression error given a finite number of samples. Unlike the initiator approximation, this approach does not introduce an additional bias. Another FCIQMC extension that can be applied to FCI-FRI involves
calculating perturbative corrections to the energy.\cite{Blunt2018} 

Due to the
versatility of the FRI framework, many recent FCIQMC extensions can also be
applied to FCI-FRI methods, which may yield further performance improvements. Here, we compare FCI-FRI methods only to the original FCIQMC method, without extensions, in order to (1) facilitate clarity in our presentation of the FCI-FRI methods, and (2) isolate the effects of different matrix and
vector compression schemes in our results. Future work will be devoted to incorporating these complementary extensions into FCI-FRI methods.

The remainder of this article is organized as follows. In Section
\ref{sec:methods}, we summarize the FRI framework in the context of the power
method for FCI calculations and describe the compression schemes considered in
this study. Efficient compression of the Hamiltonian matrix is accomplished
using a hierarchical scheme introduced in Section \ref{sec:hierMat} and discussed in more detail in Appendix \ref{sec:matFact}. In Section \ref{sec:results}, we discuss results obtained by applying these methods to five small molecular systems and compare
their statistical efficiencies. In Section \ref{sec:concl}, we summarize our key
findings and comment further on the differences among the methods in relation to
potential future research directions.

\section{Methods}
\label{sec:methods}
\subsection{The Power Method for Full Configuration Interaction Calculations}
The FCI formalism casts the treatment of a system of interacting fermions in terms of linear algebra~\cite{Knowles1984}. In the FCI-FRI and FCIQMC methods discussed here, a randomization of the power method is used to calculate observables associated with the ground-state (lowest-energy) eigenvector of the FCI Hamiltonian matrix, $\mathbf{H}$. This matrix is expressed in a Slater determinant basis for $N$ electrons in $M$ orbitals. 
Its only nonzero off-diagonal elements are those
corresponding to single and double excitations between pairs of Slater
determinants. The matrix element corresponding to a single excitation from
determinant $\ket{{K}}$ to $\ket{{L}} = \hat{c}^\dagger_a \hat{c}_i \ket{{K}}$
is
\begin{equation}
H_{LK} \equiv H_K(i \to a) = \mel{L}{ \hat{H} }{ K} = \gamma^{K}_{ia} \left(h_{ia} + \sum_{j \in \text{occ}} \mel{ i j}{}{ a j } \right)
\end{equation}
where $h_{ia}$ represents a matrix element of the one-electron component of the
Hamiltonian and $\mel{ i j }{}{ a
j }$ is an antisymmetrized two-electron repulsion integral. These are both
readily obtained from the output of a Hartree-Fock calculation. The
parity of the excitation $\gamma^K_{ia}$
is determined by the order of the orbitals
comprising the Slater determinants in this basis \cite{Holmes2016}. The sum is
over the orbitals occupied in $\ket{{K}}$. The notation $H_K(i \to a)$ will be
used throughout this paper to denote the index of an excitation from determinant
$\ket{K}$. The matrix element for the double excitation to $\ket{{M}} =
\hat{c}^\dagger_a \hat{c}^\dagger_b \hat{c}_i \hat{c}_j \ket{{K}}$ is
\begin{equation}
H_{MK} \equiv H_K(ij \to ab) = \mel{{M} }{ \hat{H} }{ {K}} = \gamma^{K}_{ia} \gamma^{K}_{jb} \mel{ab }{}{ ij}
\end{equation}
and the diagonal matrix element associated with $\ket{{K}}$ is
\begin{equation}
H_{KK} = \mel{{K} }{ \hat{H} }{ {K}} = \sum_{j \in \text{occ}} h_{jj} + \frac{1}{2} \sum_{i,j \in \text{occ}} \mel{i j }{}{ i j}
\end{equation}
The ground-state eigenvalue of this matrix is therefore the system's electronic 
energy.

Applying the generic power method to $\mathbf{H}$ involves iteratively generating a sequence of vectors,
here referred to as iterates. Each iterate $\mathbf{v}^{(\tau)}$, where $\tau$
denotes the iteration index, is obtained by multiplying the previous iterate by
the matrix $\mathbf{P} = \mathbf{1} - \varepsilon \mathbf{H}$, where $\mathbf{1}$ is
the identity and $\varepsilon$ is a positive number that is sufficiently small to
ensure that the ground state of $\mathbf{H}$ is the dominant eigenvector of
$\mathbf{P}$. 
%Like $\mathbf{H}$, the iterates are expressed in the Slater
%determinant basis. 
The initial iterate,
$\mathbf{v}^{(0)}$, must have nonzero overlap with the ground-state eigenvector,
$\mathbf{v}_\text{GS}$. In FCI, the Hartree-Fock unit vector is usually a
suitable choice and is used in all of the calculations presented here. The iterates converge to the ground-state eigenvector up to a normalization factor,
\begin{equation}
\lim_{\tau \to \infty} \frac{\mathbf{v}^{(\tau)}}{||\mathbf{v}^{(\tau)}||} = \mathbf{v}_\text{GS}
\end{equation}
After sufficiently many iterations, convergence to the ground-state is geometric, with error decaying by a factor of
$(1-\varepsilon E_0)/(1-\varepsilon E_1)$
after each iteration. Here $E_0$ is the ground-state eigenvalue of $\mathbf{H}$, and $E_1$ is the first excited-state eigenvalue.
Alternative choices of $\mathbf{v}^{(0)}$ may be used to reduce the number of iterations required for convergence~\cite{Blunt2015}.  
The norms of the
iterates $||\mathbf{v}^{(\tau)}||$ tend to either 0 or $\infty$, depending on
the sign of $E_0$, as $\tau \to \infty$. An energy shift,
$S^{(\tau)}$, is therefore included in the matrix $\mathbf{P}^{(\tau)}$ at each
iteration to stabilize the norm,
\begin{equation}
\label{eq:Puncomp}
\mathbf{P}^{(\tau)} = \mathbf{1} - \varepsilon \left( \mathbf{H} - S^{(\tau)} \mathbf{1} \right)
\end{equation}
where $S^{(\tau)}$ is updated dynamically after every $A$ iterations, where $A$ is a user-specified parameter (10 in our calculations), according
to the formula introduced in the FCIQMC method, \cite{Booth2009}
\begin{equation}
\label{eq:enShift}
S^{(\tau)} = S^{(\tau - A)} - \frac{\xi}{A \varepsilon} \ln \frac{|| \mathbf{v}^{(\tau)} ||_1}{|| \mathbf{v}^{(\tau - A)} ||_1}
\end{equation}
Here $\xi$ is a user-specified damping parameter (taken to be 0.05 in the
calculations presented here), and $|| \cdot ||_1$ denotes the one-norm, defined
for an arbitrary vector $\mathbf{x}$ as
\begin{equation}
\label{eq:oneNorm}
|| \mathbf{x} ||_1 = \sum_i |x_i|
\end{equation}
This procedure is used to stabilize the one-norm of the iterates in all methods considered in this study. In FCIQMC, the shift is updated only after the one-norm of the iterates (i.e.
the number of walkers) has reached a specified target. \cite{Booth2009} The
iterates are generated by the relation
\begin{equation}
\label{eq:friMarkov}
\mathbf{v}^{(\tau + 1)} = \mathbf{P}^{(\tau)} \mathbf{v}^{(\tau)}
\end{equation}

\subsection{FRI Compression Schemes}
\label{sec:friComp}
The size of the FCI basis, $N_{\mathrm{FCI}} \sim O(M\ \mathrm{choose}\ N)$,
renders it impossible to apply the power method as described above to many
systems of chemical interest. The memory cost is $O(N_\mathrm{FCI})$ and the
computational cost of matrix-vector multiplication is $O(N^2 V^2
N_\mathrm{FCI})$, where $V = M-N$ is the number of virtual (unoccupied)
orbitals.  For large systems, these costs are prohibitive. 
The FCI-FRI methods circumvent these bottlenecks by
stochastically compressing the vector
$\mathbf{v}^{(\tau)}$, and possibly the matrix $\mathbf{P}^{(\tau)}$, in each iteration.  Stochastic compression is defined such
that (1) the resulting compressed vector or matrix has at most a desired number
$m$ of nonzero elements and (2) the expectation value of each element in the
compressed vector or matrix is equal to the corresponding element in the input
vector or matrix, i.e.
\begin{equation}
\label{eq:compDef}
\text{E} \left[ \Phi \left(\mathbf{x} \right) \right]_i = x_i
\end{equation}
where $\Phi$ denotes the compression operation and $\mathbf{x}$ is an arbitrary
vector. The fact that many of the  elements in the compressed matrix or vector
are zero facilitates the use of sparse linear algebra schemes, which enables the
efficiency of FRI methods. 

As an example, in an FCI-FRI method that uses only vector compression, matrix-vector multiplication is performed as 
\begin{equation}
\mathbf{v}^{(\tau + 1)} 
    = \Phi\left( \mathbf{P}^{(\tau)} \mathbf{v}^{(\tau)}\right)
\end{equation}
This method has a memory cost of $O(N^2V^2 m)$ (to store the nonzero elements in the matrix-vector product before compression) and a computational cost of $O(N^2 V^2 m \log m)$. For many systems of chemical interest,
these costs can be significantly less than those for deterministic FCI.

There are many possible compression methods in FRI with the above defining properties that differ in the degree of statistical error they introduce. In order to emphasize the generality of the FRI framework, we begin by introducing several such methods in more abstract linear algebra terms before discussing their specific application to the FCI problem.

\subsubsection{Vector Compression}
\label{sec:vecComp}
In this study, we compare several different approaches to vector compression. These have been applied in previous stochastic quantum chemistry calculations, although they can be applied more generally to any vector. The simplest approach to compressing an arbitrary vector $\mathbf{x}$ involves
randomly selecting a subset of its elements, each with probability
\begin{equation}
p_i = \frac{|x_i|}{|| \mathbf{x} ||_1}
\end{equation}
The expected number of times each element is sampled is
\begin{equation}
\text{E}[n_i] = mp_i
\end{equation}
where $m$ is the total number of elements selected. Therefore, assigning each element of the compressed vector the value
\begin{equation}
\Phi(\mathbf{x})_i = \frac{n_i ||\mathbf{x}||_1 \text{sgn}(x_i)}{m}
\end{equation}
ensures that the condition in eq \ref{eq:compDef} is satisfied and that the vector has at most $m$ elements (fewer if any $n_i > 1$). Possible methods for randomly generating the values $\lbrace n_i \rbrace$ will be discussed below.

It is often beneficial to preserve the largest-magnitude elements of
$\mathbf{x}$ exactly in order to reduce the overall statistical error incurred
in compressing the vector. Lim and Weare\cite{Lim2017} proposed the following
criterion for determining the number $\rho$ to preserve exactly. If $\mathbf{s}$
is a vector, with length $\ell$, of indices that sorts the elements of $\mathbf{x}$ in order of
decreasing magnitude (i.e. $|x_{s_j}| \geq |x_{s_{j+1}}|$ for  all $j < \ell$),
then $\rho$ is the minimum value of $h$ for which
\begin{equation}
\label{eq:rhoCriterion}
(m  -  h)  |x_{s_{h+1}}| \leq  \sum_{j=h+1}^c  |x_{s_j}|
\end{equation}
where $m$ denotes the desired number of nonzero elements in $\Phi(\mathbf{x})$,
and $c$ is the number of nonzero elements in $\mathbf{x}$. Thus, $\rho$ depends both on $m$ and $\mathbf{x}$. Calculating $\rho$ requires identifying the largest-magnitude elements of $\mathbf{x}$. This can be done efficiently, in $O(\rho \log c)$ time, by using a binary heap structure rather than sorting the entire vector. The  elements of
$\mathbf{x}$ with indices $\lbrace s_1, s_2, ..., s_\rho \rbrace$ are unchanged
in the compression. If $m \geq c$, this criterion naturally specifies that all
elements are preserved exactly. Otherwise, the remaining elements of
$\Phi(\mathbf{x})$ are determined by applying random sampling with $(m - \rho)$
samples to the vector $\mathbf{x}'$, which is obtained by zeroing the $\rho$
largest-magnitude elements of $\mathbf{x}$. The resulting elements of the
compressed vector are
\begin{equation}
\label{eq:CompVecEl}
\Phi(\mathbf{x})_{s_i} = \begin{cases} 
x_{s_i} & i \leq \rho \\
n_{s_i} ||\mathbf{x}'||_1 \text{sgn}(x_{s_i}) (m - \rho)^{-1} & i > \rho.
\end{cases}
\end{equation}

An alternative, deterministic approach to vector compression is preserving the $m$
largest-magnitude elements of $\mathbf{x}$ exactly and zeroing the remaining
elements. The additional sampling step introduced above has the notable advantage
that the compressed vector is equal to the original in expectation. Even with a
high degree of vector sparsity, results that are exact to within a controllable
statistical error can be obtained by averaging over many independent vector
compressions, provided there are no other sources of error.

%In the context of matrix compression, which will be described below, it will be important to be able to compress a vector without enumerating all of its nonzero elements. In such a case, efficient compression of $\mathbf{x}$ can be realized using an auxiliary vector $\mathbf{q}$ that can be compressed more efficiently. Elements of the compressed vector $\Phi_q(\mathbf{x})$ are given as
%\begin{equation}
%\label{eq:vecQ}
%\Phi_q(\mathbf{x})_i = 
%\begin{cases}
%\Phi(\mathbf{q})_i x_i / q_i & q_i \neq 0 \\
%0 & q_i = 0
%\end{cases}
%\end{equation}
%where $\Phi_q$ denotes compression using an auxiliary vector.
%Since E$[\Phi(\mathbf{q})_i] = q_i$, elements of $\Phi_q(\mathbf{x})$ are also equal to those of $\mathbf{x}$ in expectation. Any vector $\mathbf{q}$ can be used in principle, provided that it is nonzero at every position where $\mathbf{x}$ is nonzero. However, the values of elements in $\mathbf{q}$ affect the statistical error in the compressed matrix, regardless of which compression scheme is used. Choosing $\mathbf{q} = \mathbf{x}$ yields the least statistical error.

\begin{figure}
\includegraphics[width=\linewidth]{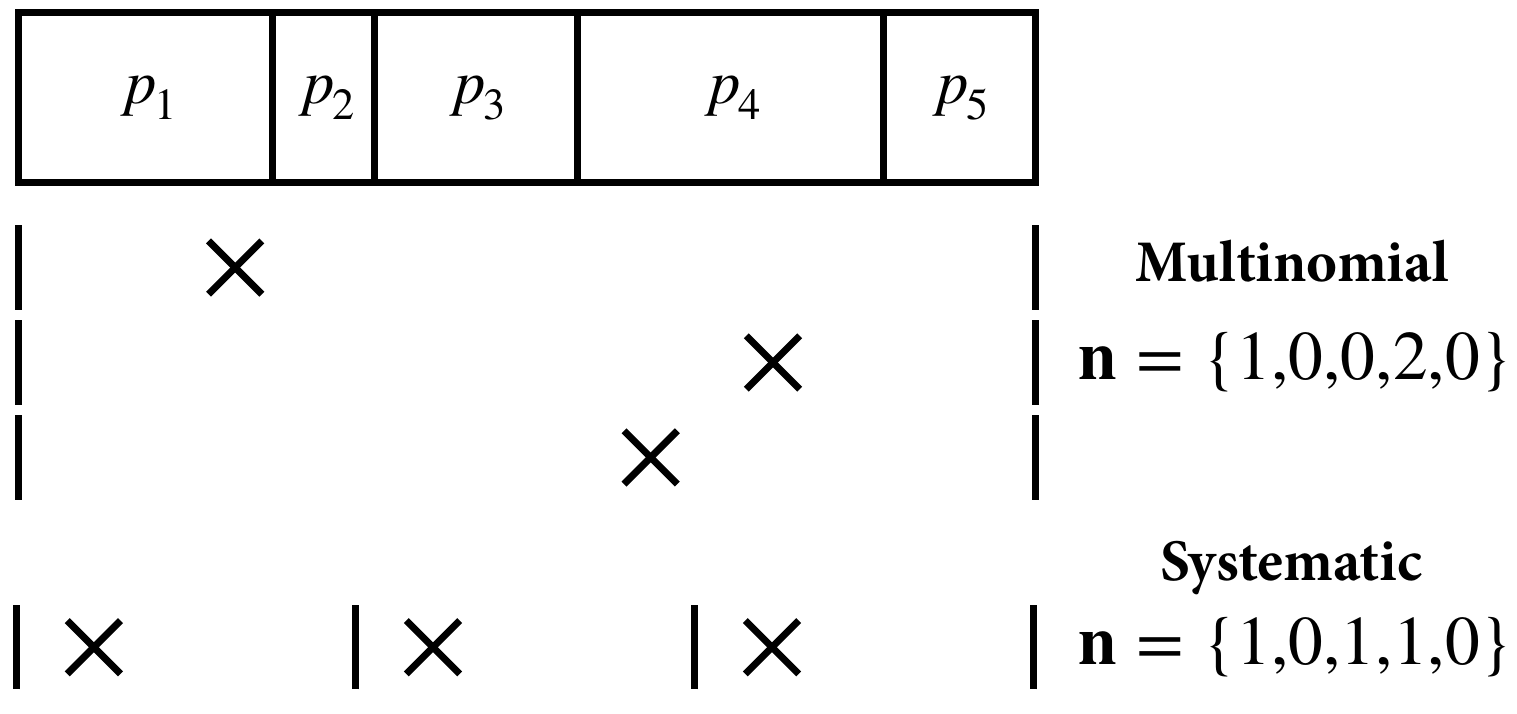}
\caption{An illustration of the multinomial and systematic sampling schemes
applied to the selection of $m=3$ elements from a probability distribution $\mathbf{p}$. The  $\times$'s represent the random numbers  $U_k$  generated on the
interval $(0,1)$. The indices selected in both schemes correspond to the
intervals in $\mathbf{p}$ with which the $\times$'s are  aligned. The vector
$\mathbf{n}$ shown for each  scheme represents the number of times each element
is selected.}
\label{fig:multiSys}
\end{figure}

\subsubsection{Sampling Schemes}
\label{sec:samplScheme}
We compare two approaches to generating the integers $\lbrace n_i \rbrace$ used for vector compression in eq \ref{eq:CompVecEl}. Both involve selecting $m$ (or $m-\rho$) elements from a probability
distribution $\mathbf{p}$ and are summarized in Figure~\ref{fig:multiSys}. In
\textit{multinomial} sampling, selections are made independently. The simplest
implementation involves generating $m$ random numbers $\lbrace U_k \rbrace$
uniformly on the interval $(0,  1)$. The index of the $k^\text{th}$ element
selected is the value of  $j$ which satisfies
\begin{equation}
\sum_{i=1}^{j-1} p_i \leq U_k < \sum_{i=1}^j p_i
\end{equation} 
Any index  can  potentially be selected  more than once, as the random numbers
$\lbrace U_k \rbrace$  are  generated  independently. 
The alias method is a more efficient
implementation of multinomial sampling than the one described
above\cite{Walker1974, Holmes2016}.

The \textit{systematic} sampling scheme typically achieves reduced variance in
the vector $\mathbf{n}$. The $m$
random numbers $\lbrace U_k \rbrace$ used in the selection of elements are
generated from a single random number $r$ chosen uniformly on the interval
$(0,1)$, as follows:
\begin{equation}
\label{eq:sysRNs}
U^{(k)} = \frac{k-1+r}{m}
\end{equation}
with $k = 1, 2, ..., m$. The  value of $r$  determines the position of the
$\times$'s in each of the $m$ subintervals of $(0,1)$ in the Systematic portion
of Figure~\ref{fig:multiSys}. The indices of elements selected are determined as
described in multinomial sampling. Although systematic sampling is expected to yield less statistical error than multinomial in general, this difference is expected to become smaller as the number of elements selected $(m)$ decreases relative to the size of the vector. When $m=1$, systematic sampling coincides exactly with multinomial sampling.

\subsubsection{Hierarchical Matrix Factorization}
\label{sec:hierMat}
The vector compression methods discussed above enable the application of FRI to iterative linear algebra methods based on matrix-vector multiplication at less cost than their deterministic counterparts. However, even the cost of multiplying a sparse vector by $\mathbf{P}^{(\tau)}$ is prohibitive for large problems in quantum chemistry. This cost can be further reduced by compressing both the matrix and vector in each iteration. In principle, the vector compression methods described above could also be applied to compress the matrix before multiplication in each iteration, e.g. by treating each of its columns as a vector. This would require enumerating all of its nonzero elements, which offers few advantages over calculating the matrix-vector product without compression.

This section describes an alternative hierarchical approach to randomly approximating a matrix-vector product using compression. For a generic matrix-vector product $\mathbf{Ax}$, this involves factoring $\mathbf{A}$ into a product of matrices and performing a sequence of vector compressions. For example, if $\mathbf{A} = \mathbf{A}^{(3)} \mathbf{A}^{(2)} \mathbf{A}^{(1)}$, then $\mathbf{Ax}$ can be approximated as:
\begin{equation}
\mathbf{Ax} = \mathbf{A}^{(3)} \Phi(\mathbf{A}^{(2)} \mathbf{x}^{(1)})
\end{equation}
where
\begin{equation}
\mathbf{x}^{(1)} = \Phi(\mathbf{A}^{(1)} \mathbf{x})
\end{equation}
The compressions after each multiplication are performed independently in this study, but other approaches in which they are not independent are possible as well.
%in some approaches like the one commonly used in FCIQMC, they are not independent. 
If $\mathbf{A}^{(1)}$, $\mathbf{A}^{(2)}$, and $\mathbf{A}^{(3)}$ are sparse, this approach can be made more efficient than calculating $\mathbf{Ax}$ directly. The multinomial selection of excitations in FCIQMC~\cite{Booth2014} can be understood as a specific implementation of this approach, but we describe it in more general terms to demonstrate that it can be used with any compression scheme in FRI.

There are multiple ways to factor the Hamiltonian matrix and correspondingly the matrix $\mathbf{P}^{(\tau)}$ for quantum chemistry calculations. These can be applied in contexts other than FCI, e.g. for stochastic coupled-cluster~\cite{Thom2010, Scott2017}. Here we consider two such factorings, near-uniform~\cite{Booth2014} and heat-bath Power-Pitzer (HB-PP)~\cite{Holmes2016, Neufeld2019}. The structure of each matrix in these factorizations is dictated by the two-body structure of the Hamiltonian. Both have the form $\mathbf{B} \mathbf{C}^{(\tau)} \mathbf{Q}$, where $\mathbf{Q}$ is factored further into a product of matrices. Elements of these matrices can be calculated efficiently using information about the symmetry of the system and, in the case of the HB-PP factorization, information from the Hamiltonian matrix. Elements of $\mathbf{Q}$ have been introduced as the probabilities for sampling excitations in previous descriptions of FCIQMC, and multiplication by $\mathbf{B}$ sums contributions from different excitations to the same determinant. 
Off-diagonal elements of the matrix $\mathbf{BQ}$ can be interpreted as an approximation to those of $\mathbf{P}^{(\tau)}$ or $\mathbf{H}$. The extra factor of $\mathbf{C}^{(\tau)}$ corrects for this discrepancy between $\mathbf{BQ}$ and $\mathbf{P}^{(\tau)}$ by multiplying by elements of $\mathbf{P}^{(\tau)}$ and dividing by elements of $\mathbf{Q}$. This form ensures that matrix elements can be calculated efficiently and that multiplication by the matrix factors is equivalent to multiplication by $\mathbf{P}^{(\tau)}$.
%The row space of each matrix, except for the final matrix, can be divided into unique subspaces, each corresponding to excitations from a single Slater determinant element in the FCI space. For example, one could consider the space containing all single excitations from $\ket{K}$, each of which is indexed by the multi-index $\lbrace K, i, a \rbrace$, where $i$ and $a$ represent occupied and virtual orbitals, respectively. The final matrix serves to map equivalent excitations from different subspaces to the same basis element. For example, if $c^\dagger_a c_i \ket{K} = c^\dagger_b c_j \ket{J} = \ket{L}$, then $\lbrace K, i, a \rbrace$ and $\lbrace J, j, b \rbrace$ are both mapped to $\ket{L}$. Additionally, one element in the $\ket{K}$ subspace, termed a ``no excitation'' element and denoted $\lbrace K, 0 \rbrace$, corresponds to the diagonal element $\mathbf{P}^{(\tau)}_{KK}$ and is mapped back to $\ket{K}$ in the final multiplication. 
The detailed forms of these factorizations are given in Appendix \ref{sec:matFact}. 

\subsection{FCI-FRI Methods Considered in this Study}
\label{sec:FRIforFCI}
The previous sections discussed compression techniques applicable to matrices and vectors in general. This section summarizes the particular implementations of these schemes in the three FCI-FRI methods considered in this study, as well as FCIQMC. A Python/Cython implementation of these methods with OpenMP parallelism is available on GitHub.\cite{resipy}

In all three FCI-FRI methods, iterate vectors are compressed systematically following matrix multiplication, regardless of which matrix compression scheme is used. A subset of $\rho$ vector elements is preserved exactly, with $\rho$ calculated as described in the discussion surrounding eq \ref{eq:rhoCriterion}, and $(m - \rho)$ additional nonzero vector elements are sampled randomly using the systematic scheme described in Section \ref{sec:samplScheme}. In order to quantify the error introduced by compressing the matrix $\mathbf{P}^{(\tau)}$ in each iteration, we considered three different matrix compression schemes in the three FCI-FRI methods. In the ``full-matrix FCI-FRI'' method, the matrix is not compressed. This method has been discussed previously and compared to FCIQMC~\cite{Lu2017}. As discussed above, its memory and CPU cost per iteration is approximately $O(N^2V^2m \log m)$. In the remaining two FCI-FRI methods, $\mathbf{P}^{(\tau)}$ is compressed either multinomially or systematically using a hierarchical factorization scheme, with additional constraints as discussed in Appendix \ref{sec:FCIcomp}. Excluding the diagonal elements of $\mathbf{P}^{(\tau)}$, which are preserved exactly, $N_\text{mat}$ samples are used in each compression. Matrix compression in ``multinomial FCI-FRI'' corresponds more closely to the scheme used in the original FCIQMC method, whereas ``systematic FCI-FRI'' is designed to reduce statistical error. These algorithms are summarized in Table \ref{tab:steps}.

\begin{table}
\caption{An overview of the steps in each iteration of the FCI-FRI methods considered in this study. The right column indicates the approximate scaling of the CPU cost of each step. The variable $N$ is the number of electrons in the system; $M$ is the number of spatial orbitals in the single-particle basis; $V = M - N$ is the number of virtual orbitals; $m$ is the number of nonzero elements kept in the solution vector; $N_\text{mat}$ is the number of off-diagonal elements sampled from the Hamiltonian matrix.}
\begin{tabular}{l | l}
\textbf{Full-matrix FCI-FRI} & CPU cost/iteration \\ \hline
1. Calculate $\mathbf{v}^{(\tau + 1)\prime} = \mathbf{P}^{(\tau)} \mathbf{v}^{(\tau)}$ & $O(N^2 V^2 m \log m)^a$ \\
2. Compress $\mathbf{v}^{(\tau + 1)\prime}$ systematically to & $O(N^2 V^2 m)$ \\
$m$ nonzero elements \\
3. Adjust the energy shift, $S^{(\tau)}$ (eq \ref{eq:enShift}) & $O(1)$
\end{tabular}
\\~\\~\\
\begin{tabular}{l | l}
\textbf{Multinomial \& systematic FCI-FRI} & CPU cost/iteration \\ \hline
1. Calculate $\mathbf{v}^{(\tau + 1)\prime} = \mathbf{P}^{(\tau)} \mathbf{v}^{(\tau)}$ using & $O(N_\text{mat})$ or $O(M N_\text{mat})$ \\
hierarchical factorization with & $+ O(N_\text{mat} \log m)^b$ \\
multinomial or systematic compression \\
to $N_\text{mat}$ nonzero elements  \\ 
2. Compress $\mathbf{v}^{(\tau + 1)\prime}$ systematically to $m$ & $O((N_\text{mat} + m) \log (N_\text{mat} + m))^c$ \\
nonzero elements \\
3. Adjust the energy shift, $S^{(\tau)}$ (eq \ref{eq:enShift}) & $O(1)$
\end{tabular}
\\~\\ \raggedright
$^a$The $(\log m)$ factor here arises because our implementation uses a less efficient binary search algorithm to perform matrix-vector multiplication. This cost could be reduced by using a hashing algorithm~\cite{Booth2014}. \\
$^bO(N_\text{mat})$ is the approximate cost of compressing the near-uniform distribution, and $O(M N_\text{mat})$ is the cost for HB-PP. The $O(N_\text{mat} \log m)$ term comes from multiplication by $\mathbf{B}$ in both factorizations and can be reduced to $O(N_\text{mat})$ using hashing. \\
$^c$Worst-case scaling. More typical scaling, corresponding to preserving relatively few elements exactly, is $O(N_\text{mat} + m)$.
\label{tab:steps}
\end{table}

\subsection{Comparison with FCIQMC}
As discussed above, the FCIQMC method described in ref \citenum{Booth2009} can be viewed as a specific method within the FRI framework. Although our presentation of the method differs somewhat from previous studies, we implemented FCIQMC in its original form, i.e. without any of its existing extensions (e.g. initiator or semi-stochastic), for comparison to FCI-FRI. This section summarizes the compression techniques in FCIQMC using the unifying language of the FRI framework, in order to facilitate comparison to the new FCI-FRI methods in this study. Further details about compression in FCIQMC can be found in Appendix \ref{sec:FCIcomp}.

In the original FCIQMC algorithm, each iterate $\mathbf{v}$ is represented by a number of signed walkers, so each of its elements $v_K$ is an integer. The total number of walkers is $||\mathbf{v}||_1$. The random selection of excitations in FCIQMC corresponds to multinomial compression of $\mathbf{P}^{(\tau)}$ using one of the factorizations discussed in Appendix \ref{sec:matFact}. The ``spawning'' step corresponds to integerization of off-diagonal elements after multiplication by $\mathbf{C}^{(\tau)}$ in the hierarchical scheme, and the ``death/cloning'' step corresponds to integerization of diagonal elements. ``Annihilation,'' i.e. the summation of matrix elements corresponding to the same Slater determinant basis element, is performed by multiplying by $\mathbf{B}$ in the hierarchical scheme.

The key difference between the original FCIQMC algorithm and multinomial FCI-FRI methods lies in the compressions performed after the final two matrix multiplications performed in the hierarchical scheme. In FCIQMC, after multiplication by $\mathbf{C}^{(\tau)}$, elements are rounded to integers using a random binomial integerization procedure. Like other vector compression techniques, this ensures sparsity in the resulting vector since many elements are rounded to zero. This reduces the cost of multiplication by $\mathbf{B}$ (i.e. ``annihilation''), since this involves summing fewer nonzero elements, but it also introduces additional statistical error. The vector obtained after multiplication by $\mathbf{B}$ is not compressed and is instead treated as the next iterate. In multinomial FCI-FRI, the vector obtained after multiplication by $\mathbf{C}^{(\tau)}$ is not compressed, so the elements that are summed during multiplication by $\mathbf{B}$ are real-valued (i.e. not necessarily integers). Sparsity is instead enforced by compressing the iterate systematically after the final matrix multiplication. It should be noted that compression is performed after multiplication by $\mathbf{B}$ in the semi-stochastic FCIQMC extension, as in FCI-FRI, although this extension was not considered in this study.

One advantage of FCIQMC is its straightforward parallelizability. Since elements are selected independently in the multinomial matrix compression scheme, they can be selected in parallel. Similarly, the stochastic rounding of matrix elements to integers can be performed in parallel, as each element is treated independently. In contrast, elements are not selected independently in systematic compression, so these strategies cannot be applied in exactly the same way. Nevertheless, parallelizing systematic schemes is possible, e.g. by performing parallel compressions in subspaces of the Slater determinant space. Investigation of these strategies will be the subject of future research. The original FCIQMC method and FCI-FRI methods become more similar as the number of nonzero elements in the compressions (number of walkers) decreases relative to the size of the basis $(N_\text{FCI})$: the probability of choosing repeated elements in multinomial matrix compression decreases, and the frequency of annihilation events in FCIQMC decreases. However, our examples suggest that the number of walkers required to obtain reasonable results from the original FCIQMC method is already sufficient to observe a substantial benefit from FRI.

\begin{table*}
\caption{The parameters used in calculations on each of the systems in this study. Unless otherwise specified, the geometry is the diatomic bond length. MP2 natural orbitals with occupancies below the occupancy threshold, if specified, were excluded from the single-particle basis. The resulting number of (spatial) orbitals is reported as $M$. The  number of unfrozen electrons considered for each system is $N$, and $N_\text{FCI}$ is the size of the FCI basis. The parameter $\varepsilon$ (eq \ref{eq:Puncomp}) is chosen to ensure convergence of the power method. $E_\text{FCI}$ denotes the exact FCI energy (including nuclear repulsion) used for comparison to our stochastic results.}
\begin{tabular}{l | c | c | c | c | c | c}
& & Occupation & & & & \\
System & Geometry & threshold / $10^{-4}$ & ($N, M)$ & $N_\text{FCI} / 10^6$ & $\varepsilon/10^{-4} E_h$ & $E_\text{FCI} / E_h$ \\ \hline
Ne (aug-cc-pVDZ) & - & - & (8, 22) & 6.69 & 10 &  $-128.709476^\text{a}$ \\
HF (cc-pCVDZ) & $0.91622$ \AA & - & (10, 23) & 283 & 1 & $-100.270929^\text{b}$\\
\ce{H2O} (cc-pVDZ) & $r_{\text{O} - \text{H}}  = 0.975512$ \AA & 6 & (10, 18) & 18.3 & 10 & $-76.167449^\text{b}$ \\
& $\angle_\text{HOH} = 110.565^\circ$ & & & & &\\
\ce{N2} (cc-pVDZ) & $1.0944$ \AA & 30 & (10, 17) & 4.8 & 5 & $-109.228042^\text{b}$ \\
\ce{C2} (cc-pVDZ) & $1.27273$ \AA & 5 & (8, 22) & 6.7 & 5 & $-75.7260112^\text{b}$ 
\end{tabular}
\begin{flushleft}
\textsuperscript{a}From ref \citenum{Olsen1996} \\
\textsuperscript{b}Calculated using the PySCF software package \cite{Sun2018}
\end{flushleft}
\label{tab:params}
\end{table*}

\subsection{Statistical Error Analysis}
\label{sec:errors}
Although in principle the iterates can be averaged to obtain an estimate of the ground-state eigenvector, the memory requirements of such an approach are prohibitive for large systems. In practice, we are only interested in observables calculated from the ground-state eigenvector, so their average values are accumulated rather than the eigenvector itself. This section addresses the calculation of the average ground-state energy and the methods used to quantify the statistical error in this average.

Conventionally, the energy of a state vector $\mathbf{x}$ is calculated as a Rayleigh quotient, defined here as:
\begin{equation}
\label{eq:rayEn}
E_\mathrm{R} (\mathbf{x}) = \frac{\mathbf{x}^* \mathbf{H} \mathbf{x}}{\mathbf{x}^* \mathbf{x}}
\end{equation}
where $\mathbf{x}^*$ denotes the conjugate transpose of $\mathbf{x}$. Averages of the energy obtained from the Rayleigh quotient estimator applied to an ensemble of random vectors will exhibit a statistical bias due to the products of correlated random vectors in both the numerator and denominator.\cite{Overy2014} Consequently, a projected energy estimator is instead used to calculate averages:
\begin{equation}
\label{eq:projEst}
E_\text{P}(\mathbf{x}) = \frac{\mathbf{v}_\text{ref}^* \mathbf{H} \mathbf{x}}{\mathbf{v}_\text{ref}^* \mathbf{x}}
\end{equation}
where $\mathbf{v}_\text{ref}$ is a constant, appropriately chosen reference vector. In principle, using a reference vector that is closer to the exact ground-state eigenvector of the Hamiltonian will yield a better estimate of the correlation energy~\cite{Alavi2016}. In this study we use the Hartree-Fock unit vector for simplicity. If this estimator is to be applied to multiple vectors $\mathbf{x}$ (in this case, the iterates obtained after each iteration), the numerator can be calculated efficiently by storing the matrix-vector product $\mathbf{H} \mathbf{v}_\text{ref}$ and taking its inner product with each vector $\mathbf{x}$. In the FCI-FRI methods in this study, this inner product is calculated before each iterate is compressed.

The numerator and denominator of eq \ref{eq:projEst} at a particular iteration are denoted as
\begin{equation}
n^{(\tau)} = \mathbf{v}_\text{ref}^* \mathbf{H} \mathbf{v}^{(\tau)}
\end{equation}
and
\begin{equation}
d^{(\tau)} = \mathbf{v}_\text{ref}^* \mathbf{v}^{(\tau)}
\end{equation}
Because $n^{(\tau)}$ and $d^{(\tau)}$ are correlated within each iteration due to their mutual dependence on $\mathbf{v}^{(\tau)}$, averaging the quotients  $n^{(\tau)} / d^{(\tau)}$ over all iterations would introduce a statistical bias. Therefore, the mean energy is calculated instead as $\langle E_\text{P} \rangle = \langle n \rangle / \langle d \rangle$, where 
\begin{equation}
\label{eq:numAve}
\langle n \rangle = \frac{1}{N_i - \tau_c} \sum_{\tau \geq \tau_c} n^{(\tau)}
\end{equation}
and the corresponding expression for the denominator is defined analogously. Here the total number of iterations in the trajectory is denoted $N_i$, and the equilibration time, $\tau_c$, is the number of iterations at the beginning of the trajectory not included in the average. Our approach to determining $\tau_c$ will be described below.
%$\langle n \rangle$ and $\langle d \rangle$ represent averages over all iterations after an equilibration time $\tau_c$ (discussed below).
%and
%where $\langle n \rangle$ and $\langle d \rangle$ represent the averages of the numerator and denominator. 
If the expected value of the iterates $\mathbf{v}^{(\tau)}$ converges to the exact ground-state eigenvector (to within a normalization factor) after infinitely many iterations, the mean energy will also converge to its exact value, since the numerator and denominator are averaged separately. In practice, a systematic bias is still observed after infinitely many iterations in FCI-FRI and FCIQMC because the expected value of the iterates does not converge to the exact ground-state eigenvector. This has been discussed previously in the context of FCIQMC and diffusion Monte Carlo methods as the population control bias.\cite{Umrigar1993, Vigor2015}

The delta method is used to calculate the variance of the average $\langle E_p \rangle$ as follows:
\begin{equation}
\label{eq:delEp}
\begin{aligned}
\text{Var}[\langle E_\text{P} \rangle] &= \text{Var}\left[\frac{\langle n \rangle}{\langle d \rangle} - \frac{n_0}{d_0} \right] \\
&\approx \text{Var} \left[\frac{\langle n \rangle - n_0}{d_0} - \frac{n_0(\langle d \rangle - d_0)}{d_0^2} \right] \\
&= \text{Var} \left[\frac{\langle n \rangle}{d_0} - \frac{n_0 \langle d \rangle}{d_0^2} \right]
\end{aligned}
\end{equation}
where $n_0$ and $d_0$ represent the deterministic quantities $\mathbf{v}_\text{ref}^* \mathbf{H} \mathbf{v}_\text{GS}$ and $\mathbf{v}_\text{ref}^* \mathbf{v}_\text{GS}$, up to an irrelevant normalization factor. We define $E^{(\tau)}_\text{delta}$ as
\begin{equation}
\begin{aligned}
E^{(\tau)}_\text{delta} &= \frac{n^{(\tau)}}{d_0} - \frac{n_0 d^{(\tau)}}{d_0^2} \\
&\approx \frac{n^{(\tau)}}{\langle d \rangle} - \frac{\langle n \rangle d^{(\tau)}}{\langle d \rangle^2}
\end{aligned}
\end{equation}
Because subsequent iterates in a trajectory are correlated, the variance in eq \ref{eq:delEp} cannot be calculated naively as $\sigma^2/(N_i - \tau_c)$, where $\sigma^2$ is the mean squared deviation from the average, i.e.
\begin{equation}
\sigma^2 = \frac{1}{N_i - \tau_c}\sum_{\tau \geq \tau_c} \left( E^{(\tau)}_\text{delta} \right)^2
\end{equation}
Instead, $\sigma^2$ must be multiplied by the integrated autocorrelation time (IAT), a measure of the degree of correlation. The IAT is estimated using the iterative procedure described in ref \citenum{Sokal1997}, as implemented in the emcee software package~\cite{Foreman2013}, using the sequence of values $\lbrace E^{(\tau)}_\text{delta} \rbrace$ as the input. If the sequence $\lbrace n^{(\tau)}  / d^{(\tau)} \rbrace$ was used instead, the resulting variance would not correspond to an energy estimate in which the numerator and denominator are averaged separately.

The equilibration time $\tau_c$ is determined for each trajectory by inspecting plots of the  IATs of the numerator and denominator of the energy estimator separately vs. $\tau_c$. Typically, the IAT is greater for smaller values of $\tau_c$, both because of their dependence on the initial iterate $\mathbf{v}^{(0)}$ and because iterates can become trapped around metastable energy values before converging to the ground-state eigenvector~\cite{Chodera2016}. Equilibration times were therefore chosen to exclude this initial period of decreasing IATs. In FCIQMC, $\tau_c$ is also constrained to be greater than the first index at which the energy shift is updated (eq \ref{eq:enShift}).

The Flyvbjerg-Petersen blocking method\cite{Flyvbjerg1989} has been used in previous FCIQMC studies\cite{Booth2009, Spencer2012, Blunt2015, Vigor2016} to calculate the variance. The approach described here has the notable advantage that no data from after the initial equilibration period ($\tau \geq \tau_c$) is discarded in the calculation of the mean and variance. Either of these methods requires a very long trajectory to achieve an accurate estimate of the variance, and it is likely that some of the statistical error estimates reported in this study are not fully converged. 

The standard error of the energy estimator is calculated as
\begin{equation}
\sigma_e = \left( {\text{Var}[\langle E_\text{P} \rangle]} \right)^{1/2}
\end{equation}
This error is expected to scale as $(N_i - \tau_c)^{-1/2}$ after sufficiently many iterations, according to the Markov chain central limit theorem with standard assumptions of ergodicity~\cite{Chung1960, Sokal1997}. This scaling renders it impossible to directly compare the standard errors from two trajectories with different numbers of iterations. Therefore, the primary metric that will be used to compare the methods discussed here is the statistical efficiency, defined as~\cite{Holmes2016}
\begin{equation}
\label{eq:eff}
E = \frac{1}{\sigma_e^2 (N_i - \tau_c)}
\end{equation}
For two methods executed for the same number of iterations after the equilibration period, the method with the greater statistical efficiency will typically yield less variance. From an alternative perspective, in order to achieve a target standard error, the method with greater statistical efficiency can be executed for fewer iterations. For example, to achieve a standard error of $10^{-5} E_h$, a method with statistical efficiency $E$ requires $[(10^{-5} E_h)^2 E]^{-1}$ iterations after the equilibration period. In this study, we do not normalize the efficiency based on the computational cost of each iteration. Therefore, for a given FCI-FRI method applied to a particular system, increasing the number of matrix or vector samples increases the statistical efficiency due to the expected decrease in error, regardless of the corresponding increase in computational cost. For this reason, when comparing the statistical efficiencies of different FCI-FRI methods and FCIQMC, we ensure that the same number of matrix and vector samples are used in all methods for each system. This ensures that any differences in the resulting statistical efficiencies are due to features inherent to the methods.

\section{Results}
\label{sec:results}
The methods described in the previous section are applied to a subset of  the molecular systems considered in ref \citenum{Booth2009}. The parameters relevant to the Hartree-Fock and randomized FCI calculations performed for these five systems  are presented in Table  \ref{tab:params}. In order to run calculations for sufficiently many iterations to obtain  robust estimates  of the mean energy and associated standard error, fewer single-particle orbitals are used for three systems than in ref \citenum{Booth2009}, thus reducing the  size of the FCI basis $(N_\text{FCI})$. This truncation is  performed by discarding natural orbitals obtained from a second-order M\o ller-Plesset  perturbation theory (MP2) calculation with occupation numbers less than  a specified threshold. We emphasize that truncating the basis is necessary only because of inefficiencies in our implementations of these methods. Optimizing our implementations should enable  the treatment of significantly larger systems. Core electrons are frozen in Ne, \ce{C2}, and \ce{N2}, as in ref \citenum{Booth2009}. The same value of $\varepsilon$  is used  to construct the matrix $\mathbf{P}^{(\tau)}$ (eq \ref{eq:Puncomp}) used in all methods for  each system. The PySCF electronic structure software package\cite{Sun2018} is  used  to perform Hartree-Fock, MP2, and deterministic FCI calculations. In ref \citenum{Booth2009}, the average FCIQMC energy for
the hydrogen fluoride (HF) molecule was compared to 
coupled-cluster theory with perturbative triple excitations, CCSD(T). Our deterministic FCI result, calculated using PySCF, differs from the CCSD(T) result by $4.89 \times 10^{-4} E_h$, and from the FCIQMC result from ref \citenum{Booth2009} by $5.4 \times 10^{-5} E_h$, a value greater than the reported uncertainty.

\subsection{FCI-FRI without Matrix Compression}
\label{sec:friFull}
\begin{figure}
\includegraphics[scale=1]{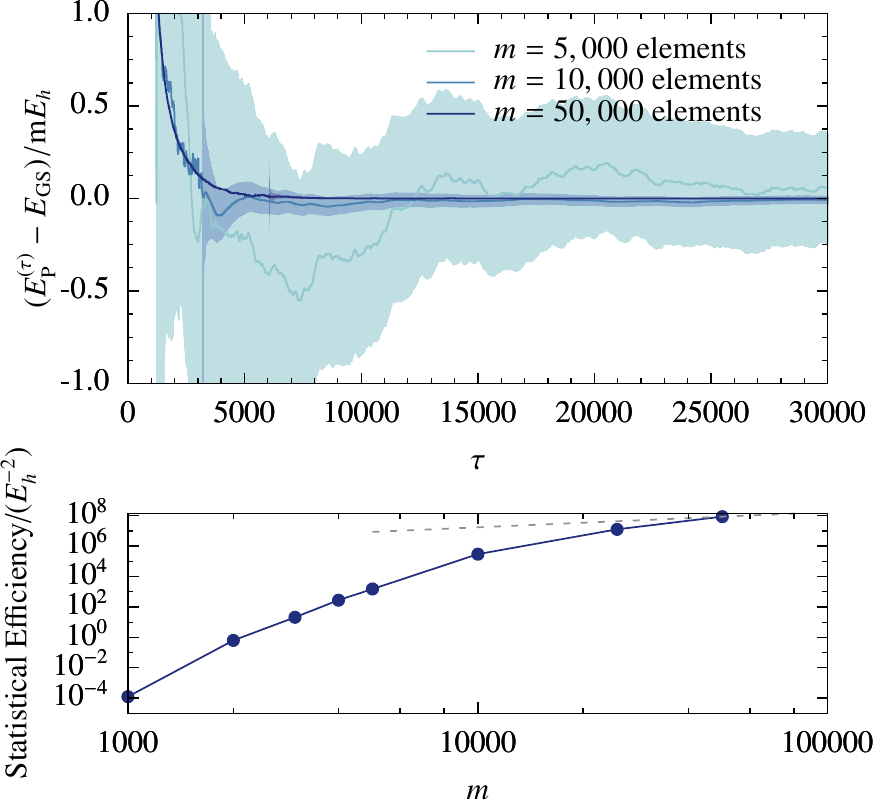}
\caption{Results obtained by applying the ``full-matrix FCI-FRI'' method to the Ne atom. (top) Differences between the energy estimator ($E_\text{P}^{(\tau)}$, eq \ref{eq:projEst}) and the exact FCI ground-state energy for three trajectories with different numbers, $m$, of nonzero elements in the compressed vectors. After the initial equilibration period $(\tau > \tau_c)$, the cumulative mean $\langle E_\text{P} \rangle$ is plotted, with the shaded region indicating the corresponding 95\% confidence interval $(\pm 2 \sigma_E)$. (bottom) The statistical efficiency for trajectories executed with different values of $m$. The dashed line with slope 1 represents the expected scaling of the efficiency with respect to $m$ (for $m$ large but less than $N_\text{FCI}$).}
\label{fig:neTrajEff}
\end{figure}

In order to isolate the contribution of vector compression to the statistical
error in calculations of the ground state energy, we first consider results obtained by
applying the ``full-matrix FCI-FRI'' method, which does not use matrix compression, to the Ne
atom. We compare calculations with differing numbers of
nonzero elements retained in the compression of each iterate ($m$). As $m$ approaches
the size of the FCI basis, this method becomes identical to the deterministic power
method. The difference between the estimated ground-state energy at each iteration and the exact energy is plotted for calculations with three different values of $m$ in the top panel of Figure~\ref{fig:neTrajEff}. The energy of the first iterate in each trajectory is the Hartree-Fock energy, since the first iterate was initialized to the Hartree-Fock unit vector. The energy decreases towards the exact energy in subsequent iterations. After the estimator is determined to be sufficiently close to the exact energy, at iteration $\tau_c$, the mean is accumulated according to eq \ref{eq:numAve}. This cumulative mean is plotted in Figure~\ref{fig:neTrajEff} for $\tau \geq \tau_c$.

The value of the equilibration time $\tau_c$ used in these trajectories increases with increasing $m$ (Table \ref{tab:friAll}), primarily due the greater degree of noise in trajectories with fewer nonzero elements in each iterate. When $m$ is smaller, the energy decreases more quickly towards the ground state, causing a lesser value of $\tau_c$, but fluctuates to a greater extent after $\tau = \tau_c$. In the deterministic power method, the asymptotic convergence rate is determined by the ratio $({1} - \varepsilon E_0)/({1} - \varepsilon E_1)$. Randomized implementations of the power method can exhibit different convergence properties, depending on the statistical error introduced in each iteration. This trend in $\tau_c$ is therefore not surprising, and it suggests that an accurate energy estimate can be achieved at less computational cost if the values of $m$ and $\varepsilon$ are varied dynamically during the calculation.

\begin{figure}
\includegraphics[scale=1]{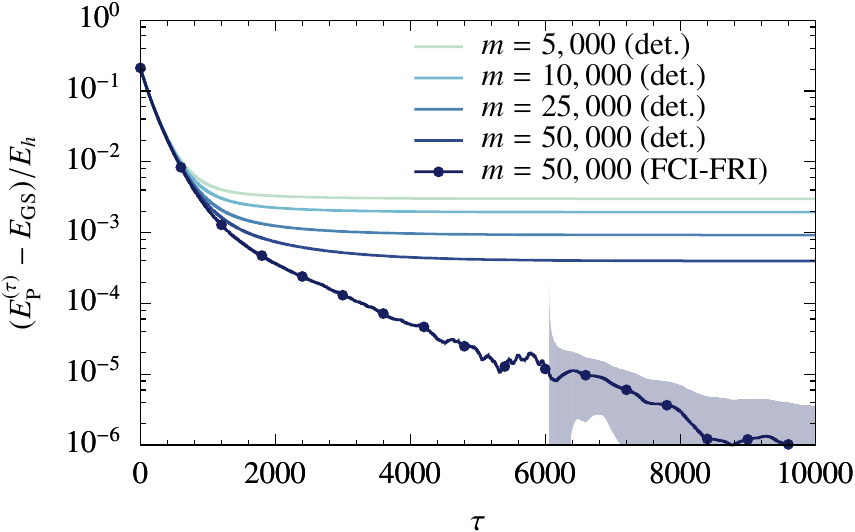}
\caption{Results obtained by applying the power method with deterministic vector truncation to the Ne atom. Only the $m$ greatest-magnitude elements of the vector were preserved exactly after each iteration. Differences between the energy estimator $E_\text{P}^{(\tau)}$ and the exact energy at each iteration are plotted for four trajectories with different values of $m$. Results from the ``full-matrix FCI-FRI'' calculation with $m=50,000$ elements from Figure \ref{fig:neTrajEff} are presented for comparison. Note the log scale on the vertical axis.
}
\label{fig:neTrajDet}
\end{figure}

The difference $E_\text{diff}$ between the final estimate of the energy, obtained by averaging over all $\tau \geq \tau_c$, and the exact FCI energy from ref \citenum{Olsen1996}, is presented for each $m$ in Table \ref{tab:friAll}. The number of iterations included in each of these averages can be obtained by subtracting $\tau_c$ from the reported total number of iterations, $N_i$. The reported uncertainties, twice the standard error $\sigma_E$ calculated as described in Section \ref{sec:errors}, represent 95\% confidence intervals for the means. The exact energy is within these confidence intervals for all values of $m$ reported here (i.e. $|E_\text{diff}| < 2 \sigma_E$). The standard error is expected to decrease after more iterations, with an asymptotic scaling of $(N_i - \tau_c)^{-1/2}$. Confidence intervals for intermediate values of $\tau$, calculated by scaling the final confidence intervals reported in Table \ref{tab:friAll}, are shown as shaded areas in Figure \ref{fig:neTrajEff}. The value $E_\text{diff}$ is not expected to converge to 0 but rather to the statistical bias, as discussed in Section \ref{sec:errors}. This bias scales as $m^{-1}$ when $m$ is sufficiently large (but still much smaller than the size of the FCI basis, $N_\text{FCI}$)\cite{Lim2017}, but the number of iterations performed in our calculations is not sufficient to measure the biases in these calculations accurately.

\begin{table}
\caption{Results obtained by applying the ``full-matrix FCI-FRI'' method to the Ne atom with different values of $m$. The difference $E_\text{diff}$ between the mean and exact (FCI) energy for each calculation is presented, with twice the standard error $\sigma_E$ (95\% confidence interval). 
%The length of the equilibration period $(\tau_c)$ and total number of iterations $(N_i)$ are reported in thousands. 
The length of the equilibration period $(\tau_c)$ and total number of iterations $(N_i)$ are given. 
The statistical efficiency is calculated using eq \ref{eq:eff}. The mean number of Hamiltonian matrix evaluations in each iteration $N_\text{mat}$ is presented for comparison to other methods.}
\begin{tabular}{c | c | c | c | c | c}
$m/10^3$ & $N_\text{mat}/10^6$ & $(E_\text{diff} \pm 2\sigma_E) / (10^{-5} E_h$) & Eff./($10^6 E_h^{-2}$) & $\tau_c/10^3$ & $N_i/10^3$  \\ \hline
1 & 0.93 & $6437 \pm 16099$ & $1.25 \times 10^{-10}$ & 0.8 & 1237 \\
2 & 1.9 & $141 \pm 242$ & $6.4 \times 10^{-7}$ & 1.1 & 1062 \\
5 & 4.7 & $-0.089 \pm 4.60$ & 0.0015 & 1.2 & 1200 \\
10 & 9.3 & $0.307 \pm 0.480$ & 0.296 & 3.2 & 589 \\
25 & 23.4 & $-0.053 \pm 0.112$ & 12.8 & 4.8 & 256 \\
50 & 46.8 & $0.034 \pm 0.063$ & 86.9 & 6.1 & 123
\end{tabular}
\label{tab:friAll}
\end{table}

In Table \ref{tab:friAll}, decreased standard error is observed in calculations with greater values of $m$, despite the fact that fewer iterations were included in these calculations. If the errors from these calculations are compared after the same number of iterations, the trend with increasing $m$ would be more pronounced. The statistical efficiency does not depend on the number of iterations and therefore allows for a more direct comparison. 
%In an approximate sense, it measures the degree of fluctuations in the energy after each iteration and the extent to which fluctuations are correlated. 
Statistical efficiencies calculated from all trajectories are presented in Table \ref{tab:friAll} and in the bottom panel of Figure~\ref{fig:neTrajEff}. While the computational cost of full-matrix FCI-FRI calculations is approximately proportional to $m$, the statistical efficiency appears to increase at a faster-than-$m$ rate for small $m$. This indicates that, in terms of reducing the standard error, it is more advantageous to increase $m$ in this pre-asymptotic regime than to increase the number of iterations. The statistical efficiency is expected to increase linearly with $m$ for $m$ sufficiently large (but still much smaller than $N_\text{FCI}$)~\cite{Lim2017}. 
Similar faster-than-$m$ pre-asymptotic scaling has been observed in other methods that use sequential Monte Carlo sampling on a classical problem~\cite{Webber2019}, suggesting that it is {not} (solely) a manifestation of the fermion sign problem in this case.

%The efficiency of full-matrix FCI-FRI calculations for Ne decreases substantially as $m$ is decreased below $5,000$ nonzero elements (Table \ref{tab:friAll}). This is likely related to the convergence behavior of the original FCIQMC method. Previous studies found that the number of walkers in a FCIQMC calculation must exceed a system-dependent critical value in order to ensure reliable convergence to the ground-state energy~\cite{Booth2009, Spencer2012}. Our results suggest that there is an analogous critical value for $m$ for each system in the full-matrix FCI-FRI method and that it is approximately 1,000-2,000 for Ne. Interestingly, the mean value of the energy shift $S^{(\tau)}$, which has been considered as an alternate estimate of the ground-state energy in previous FCIQMC studies, does not converge to the exact ground state energy for the $m=1,000$ calculation, differing by $7.7 \pm 0.007 E_h$. Additionally, the Hartree-Fock component of many iterates in this calculation was zero. Neither of these behaviors are observed in calculations with values of $m$ greater than 1,000.

Before considering the effect of matrix compression on the statistical error, we
comment briefly on the benefits of using stochastic, rather than deterministic,
vector compression. Results for the Ne atom obtained using a deterministic
vector compression scheme are presented in Figure~\ref{fig:neTrajDet}. In each
iteration, the matrix is not compressed, the $m$ greatest-magnitude elements in
the vector are preserved exactly, and the remaining vector elements are
zeroed. For all values of $m$ considered, the energy calculated from the
projected estimator, $E_\text{P}$, converges after approximately 3000 iterations. Energies obtained from the ``full-matrix FCI-FRI'' method, with $m=50,000$ nonzero elements kept after each iteration, are also presented for comparison. The error in the corresponding deterministic calculation after a similar number of iterations is almost two orders of magnitude greater than the 95\% confidence interval in the FCI-FRI calculation. Similar results for other electronic systems were observed previously in ref \citenum{Lu2017}.
These
results indicate that the success of the FCI-FRI method in these cases cannot be
attributed to its discarding vector elements that do not contribute
significantly to the energy, as is done in the deterministic approach. The stochastic representation of these
small-magnitude elements is crucial to its success. 
This observation may be relevant to selected CI 
methods~\cite{Huron1973,Tubman2016,Zhang2016, Sharma2017, Wang2019}, which utilize a similar
greedy optimization scheme.

\subsection{Methods with Matrix Compression}
The cost of the full-matrix FCI-FRI method renders it intractable for larger systems, so we also evaluate the performance of methods that use matrix compression, including the original FCIQMC method.

\subsubsection{Near-Uniform Factorization}
Methods that utilize the
near-uniform factorization described in Appendix \ref{sec:nearUniQ} will be discussed first. In
order to ensure a fair comparison among these methods, all calculations for each
system are executed with approximately the same cost, i.e.~using the same
numbers of nonzero elements in the matrix and vector compressions
in each iteration ($N_\text{mat}$ and $m$, respectively). In an FCIQMC
calculation, $N_\text{mat}$ is the number of walkers, and $m$ is determined by
their distribution among the Slater determinant basis elements. In FCIQMC, the number of walkers and $m$
fluctuate randomly in each iteration. Previous studies have determined that the number of walkers must be greater than a system-dependent critical value in order to ensure convergence. The number of walkers used in the FCIQMC calculations discussed here are constrained to be greater than these critical values. Critical values for the Ne and HF systems are
given in ref \citenum{Booth2009}, and those for the remaining systems
considered in this study are determined using the same scheme, i.e. by observing trends
in the growth of the number of walkers before the energy shift $S^{(\tau)}$ is
updated. The values of
$N_\text{mat}$ and $m$ used in FCI-FRI calculations are fixed at the corresponding average
values obtained from the FCIQMC calculations after walker growth has stabilized.

Results from these calculations for all molecular systems are presented in Table \ref{tab:allNearUni}. In all calculations, average energies converge to the
exact FCI energies reported in Table \ref{tab:params} to within twice the standard error (95\% confidence interval). Strictly speaking, all methods considered here exhibit a statistical bias, although for these calculations it is very likely less than the reported confidence intervals. After more iterations, we expect that the standard error for all trajectories will decrease, and the energy differences $E_\text{diff}$ for both trajectories of a particular method and system will converge to the same statistically significant bias. It is impossible to draw definitive conclusions about the relative biases of the three methods described here without more iterations.

%Our results suggest that this may not be the case for FCIQMC applied to \ce{C2}. This strongly correlated system has proven challenging in other FCIQMC studies~\cite{Booth2011}, necessitating the use of a relatively large number of walkers~\cite{Booth2009}. In preliminary analyses, we applied FCIQMC to this system with only 3.10 million walkers on average, i.e. more than the critical value for this system but less than in the calculations presented in Table \ref{tab:allNearUni}. This calculation showed a greater bias (in units of $10^{-5} E_h$, $53.27 \pm 20.66$). This suggests that reducing the error in matrix and vector compression, as is done in the multinomial and systematic FRI methods, may reduce this bias and enable the use of fewer matrix and vector samples than in FCIQMC. However, given the magnitude of the standard errors in these calculations, it is impossible to draw definitive conclusions about their relative biases.

\begin{figure}
\includegraphics[scale=1]{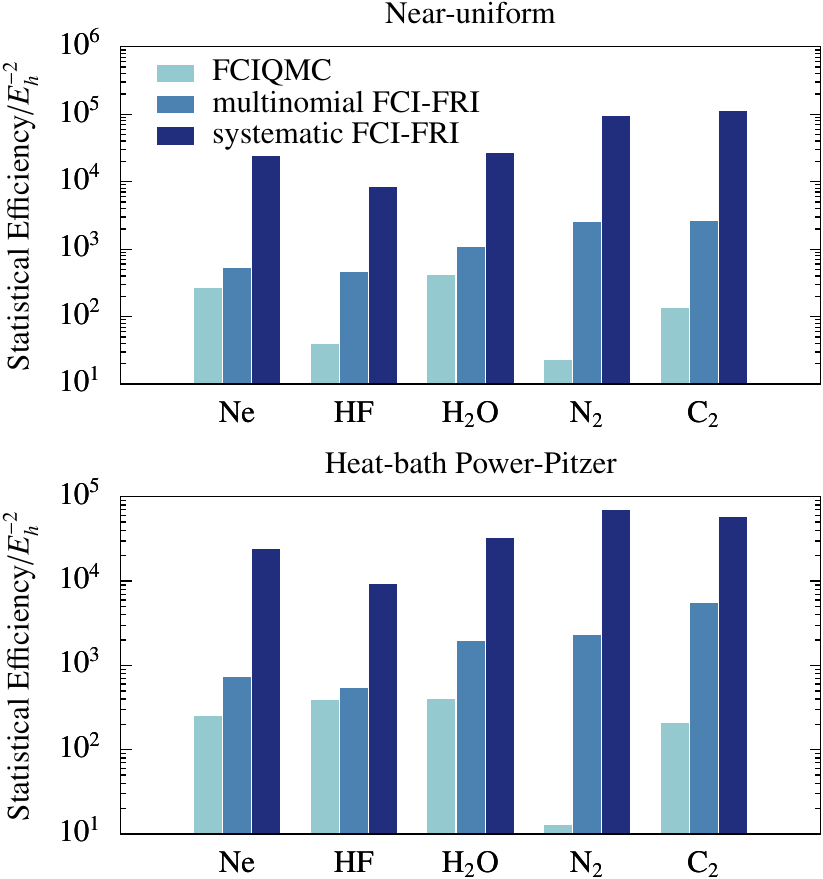}
\caption{Increases in statistical efficiency are robust across five molecular systems and two choices of matrix factorization schemes, near-uniform (top) and heat-bath Power-Pitzer (bottom). Reported statistical efficiencies represent an average over the two independent trajectories obtained using each method and do not reflect differences in computational cost for systems with different sizes. Note the log scale on the y-axis.}
\label{fig:allEff}
\end{figure}

Standard errors from FCIQMC calculations range from $3 \times 10^{-5} E_h$ to $20 \times 10^{-5} E_h$, while those from the FCI-FRI methods are smaller ($2 \times 10^{-5} E_h$ to $6 \times 10^{-5} E_h$ for multinomial FCI-FRI, and $0.4 \times 10^{-5} E_h$ to $1.7 \times 10^{-5} E_h$ for systematic FCI-FRI), \textit{despite their use of fewer iterations}. This trend is also reflected in the corresponding efficiencies (Figure~\ref{fig:allEff}, top), which are normalized based on the different number of iterations considered in the calculation of each standard error. For all systems, efficiencies for systematic FCI-FRI calculations are more than an order of magnitude greater than those for multinomial FCI-FRI calculations, which are in turn 2 to 113 times greater than those for FCIQMC calculations.

The integrated autocorrelation times (IATs), calculated as described in Section \ref{sec:errors} for all three methods, are similar within each system considered here. This is likely because the same value of the imaginary time step, $\varepsilon$, is used for each system (Table \ref{tab:params}). A previous study~\cite{Holmes2016} found that reducing the statistical error in matrix compression in FCIQMC enabled the use of greater values of $\varepsilon$. This reduces the degree of correlation between iterates, thereby decreasing the IAT and increasing the statistical efficiency. This suggests that using greater values of $\varepsilon$ in the multinomial and systematic FCI-FRI methods could potentially increase the observed difference in their efficiencies. Furthermore, increasing $\varepsilon$ may reduce the equilibration times $\tau_c$ for the FCI-FRI methods.
%To test this possibility, we execute 4 additional calculations on \ce{C2} using the multinomial and systematic FRI methods with twice the value of $\varepsilon$ reported in Table \ref{tab:params}. The efficiencies of these calculations appear to be greater than their counterparts in Table~\ref{tab:allNearUni}, although it is difficult to accurately quantify this difference without executing these calculations for more iterations. Furthermore, increasing $\varepsilon$ may reduce the equilibration times $\tau_c$ for the FRI methods.
%, which were greater for many systems than in FCIQMC. This trend of increasing $\tau_c$ with decreasing error mirrors that for the full-matrix FRI method discussed in Section \ref{sec:friFull}.

Because the systematic FCI-FRI method converges to the deterministic power method as $N_\text{mat}$ and $m$ approach finite values, we expect that the reported performance advantages for systematic FCI-FRI relative to the other two methods would increase for greater values of $N_\text{mat}$ and $m$. On the other hand, because the compression schemes used in these FCI-FRI methods become more similar to those in FCIQMC as the size of the FCI basis increases relative to $N_\text{mat}$ and $m$, the statistical efficiencies of these methods are expected to become more similar in this limit. For many systems, however, the values of $N_\text{mat}$ and $m$ required to calculate reasonably accurate energy estimates also increase with system size. In the calculations we have compared thus far, the values of these parameters are dictated by the critical number of walkers in FCIQMC~\cite{Booth2009}. Calculations for the Ne and HF systems were also compared with fewer matrix and vector samples. Using only 164,000 walkers in an FCIQMC calculation on Ne yields an energy estimate that differs from the exact energy by $(-163 \pm 20783) \times 10^{-5} E_h$, whereas a systematic FCI-FRI calculation with equivalent numbers of samples yields an energy estimate that differs by $(0.58 \pm 5.15) \times 10^{-5} E_h$ after a similar number of iterations. The efficiencies of these two calculations differ by seven orders of magnitude. A similar comparison for HF with only 812,000 walkers also shows a factor of $10^7$ difference in efficiencies. This suggests that FCI-FRI methods may allow for the use of significantly fewer matrix and vector samples than the original FCIQMC method.

%However, FCI-FRI methods may offer significant advantages over the original FCIQMC method at moderate values of $N_\text{mat}$ and $m$, before this limit is reached. The values of $N_\text{mat}$ and $m$ used in the comparisons discussed thus far are determined by the critical number of walkers from the original FCIQMC method~\cite{Booth2009}. We also compared FCIQMC and systematic FCI-FRI calculations for the Ne and HF systems with fewer walkers. 

\begin{table*}
\caption{Differences between mean energy estimates and those reported in Table \ref{tab:params} $(E_\text{diff})$ for each of the systems considered here calculated using the FCIQMC, multinomial FCI-FRI, and systematic FCI-FRI methods with the near-uniform factorization scheme. The parameter $m$ represents the sparsity of the iterates (mean sparsity for FCIQMC), and $N_\text{mat}$ represents the number of Hamiltonian matrix elements evaluated in each iteration (mean number of walkers for FCIQMC). Results from two  independent trajectories are presented for  each method. Mean energy differences $\pm$ twice the standard error (95\% confidence interval) are reported for each calculation, followed by the length of the equilibration period ($\tau_c$) and total number of iterations ($N_i$). For each chemical system, the three methods share a similar computational cost per iteration.}
\begin{tabular}{c | c | c | c | c | c | c | c | c | c | c | c}
& & & \multicolumn{3}{c |}{FCIQMC} & \multicolumn{3}{c |}{multinomial FCI-FRI} & \multicolumn{3}{c }{systematic FCI-FRI} \\
System & $m/10^3$ & $N_\text{mat}/10^6$ & ($E_\text{diff} \pm 2\sigma_E$)/($10^{-5} E_h$) & $\tau_c/10^3$ & $N_i/10^3$ & ($E_\text{diff} \pm 2\sigma_E$)/($10^{-5} E_h$) & $\tau_c/10^3$ & $N_i/10^3$ & ($E_\text{diff} \pm 2\sigma_E$)/($10^{-5} E_h$) & $\tau_c/10^3$ & $N_i/10^3$ \\ \hline
Ne & 242 & 0.26 & $-1.44 \pm 7.36$ & 22.5&2800  & $0.06 \pm 5.66$ & 15.0&2373  & $-0.16 \pm 1.09$ & 11.5&1422  \\ 
& & & $2.89 \pm 7.47$ & 22.5&2800  & $-3.12 \pm 4.99$ & 15.0&3200  & $-0.74 \pm 1.11$ & 11.0&1445  \\ \hline
HF & 926 & 1.00 & $10.57 \pm 26.86$ & 160.0&1469  & $-9.76 \pm 11.17$ & 400.0&1104  & $0.49 \pm 2.57$ & 620.0&1495  \\ 
& & & $21.09 \pm 33.50$ & 430.0&1474  & $-7.03 \pm 11.28$ & 380.0&1100  & $-0.37 \pm 3.37$ & 620.0&994  \\ \hline
\ce{H2O} & 491 & 0.57 & $-0.96 \pm 6.52$ & 30.0&2400  & $0.61 \pm 5.54$ & 20.0&1232  & $-0.41 \pm 1.29$ & 25.0&1055  \\ 
& & & $0.54 \pm 6.47$ & 30.0&2400  & $-2.08 \pm 5.63$ & 20.0&1228  & $0.17 \pm 1.16$ & 25.0&1059  \\  \hline
\ce{N2} & 1014 & 1.21 & $-7.46 \pm 29.75$ & 200.0&1788  & $-1.05 \pm 5.02$ & 80.0&822  & $0.14 \pm 0.82$ & 76.7&554  \\ 
& & & $4.78 \pm 39.85$ & 200.0&1791  & $2.41 \pm 5.55$ & 52.1&512  & $-0.89 \pm 1.33$ & 170.0&557  \\ \hline
\ce{C2} & 2622 & 4.14 & $9.53 \pm 9.56$ & 50.0&2908  & $1.32 \pm 3.55$ & 540.0&2051  & $0.71 \pm 1.08$ & 42.2&513  \\ 
& & & $4.76 \pm 11.54$ & 50.0&2768  & $-2.30 \pm 3.92$ & 450.0&1327  & $-0.50 \pm 0.77$ & 50.6&516  \\ 
\end{tabular}
\label{tab:allNearUni}
\end{table*}

\subsubsection{Heat-Bath Power-Pitzer Factorization}
Results obtained using the three methods with the HB-PP factorization matrix mostly follow the same trends as those for the near-uniform factorization (Table \ref{tab:allHeatBath}). Standard errors for systematic and multinomial FCI-FRI calculations are less than those from FCIQMC, as is reflected in their associated efficiencies (Figure \ref{fig:allEff}, bottom). One FCIQMC calculation on \ce{H2O} did not converge to within the 95\% confidence interval, although given the relative magnitude of its standard error, this is likely a statistical anomaly. Systematic FCI-FRI calculations on \ce{C2} were particularly expensive due to the number of orbitals and cost of evaluating elements of matrices in the HB-PP factorization, rendering it difficult to accumulate sufficiently many samples to obtain an accurate estimate of the integrated autocovariance. Consequently, the estimated standard errors for these calculations are likely more inaccurate than for the other calculations in this study. This highlights the need for more efficient implementations of these FCI-FRI methods.

\begin{table*}
\caption{Mean energy differences $\pm$ twice the standard error for randomized methods using the heat-bath Power-Pitzer factorization scheme. Parameters are reported for each trajectory as in Table \ref{tab:allNearUni} (iterate vector sparsity, number of matrix samples, and number of iterations).}
\begin{tabular}{c | c | c | c | c | c | c | c | c | c | c | c}
& & & \multicolumn{3}{c |}{FCIQMC} & \multicolumn{3}{c |}{multinomial FCI-FRI} & \multicolumn{3}{c }{systematic FCI-FRI} \\
System & $m/10^3$ & $N_\text{mat}/10^6$ & ($E_\text{diff} \pm 2\sigma_E$)/($10^{-5} E_h$) & $\tau_c/10^3$ & $N_i/10^3$ & ($E_\text{diff} \pm 2\sigma_E$)/($10^{-5} E_h$) & $\tau_c/10^3$ & $N_i/10^3$ & ($E_\text{diff} \pm 2\sigma_E$)/($10^{-5} E_h$) & $\tau_c/10^3$ & $N_i/10^3$ \\ \hline
Ne & 242 & 0.26 & $0.01 \pm 13.43$ & 15.0&902  & $3.96 \pm 7.83$ & 15.0&917  & $-0.44 \pm 1.61$ & 15.0&657  \\ 
& & & $-3.41 \pm 13.22$ & 20.0&963  & $0.75 \pm 7.92$ & 15.0&905  & $-1.09 \pm 1.61$ & 15.0&686  \\ \hline
HF & 926 & 1.00 & $-4.15 \pm 17.31$ & 130.0&502  & $-4.78 \pm 18.28$ & 180.0&436  & $-0.91 \pm 2.99$ & 40.0&447  \\ 
& & & $4.41 \pm 15.89$ & 120.0&507  & $0.66 \pm 13.42$ & 50.0&430  & $-0.54 \pm 3.00$ & 27.4&654  \\ \hline
\ce{H2O} & 491 & 0.57 & $-12.33 \pm 10.66$ & 30.0&938  & $-1.53 \pm 5.95$ & 20.0&645  & $-0.30 \pm 1.52$ & 20.0&533  \\ 
& & & $-4.04 \pm 10.45$ & 30.0&936  & $-3.65 \pm 5.69$ & 20.0&646  & $-0.18 \pm 1.63$ & 20.0&531  \\  \hline
\ce{N2} & 997 & 1.15 & $33.68 \pm 94.08$ & 200.0&663  & $1.03 \pm 5.53$ & 53.2&699  & $0.43 \pm 1.18$ & 64.6&373  \\ 
& & & $55.74 \pm 75.77$ & 200.0&659  & $4.19 \pm 5.07$ & 57.1&700  & $-0.19 \pm 1.72$ & 72.7&372  \\ \hline
\ce{C2} & 2620 & 4.14 & $-11.20 \pm 17.99$ & 50.0&573  & $-1.02 \pm 4.59$ & 190.0&432  & $-0.15 \pm 2.02$ & 130.0&331  \\ 
& & & $12.40 \pm 22.95$ & 140.0&581  & $-0.12 \pm 5.61$ & 36.8&494  & $-0.73 \pm 1.96$ & 50.0&213  \\ 
\end{tabular}
\label{tab:allHeatBath}
\end{table*}

\begin{figure}
\includegraphics[scale=1]{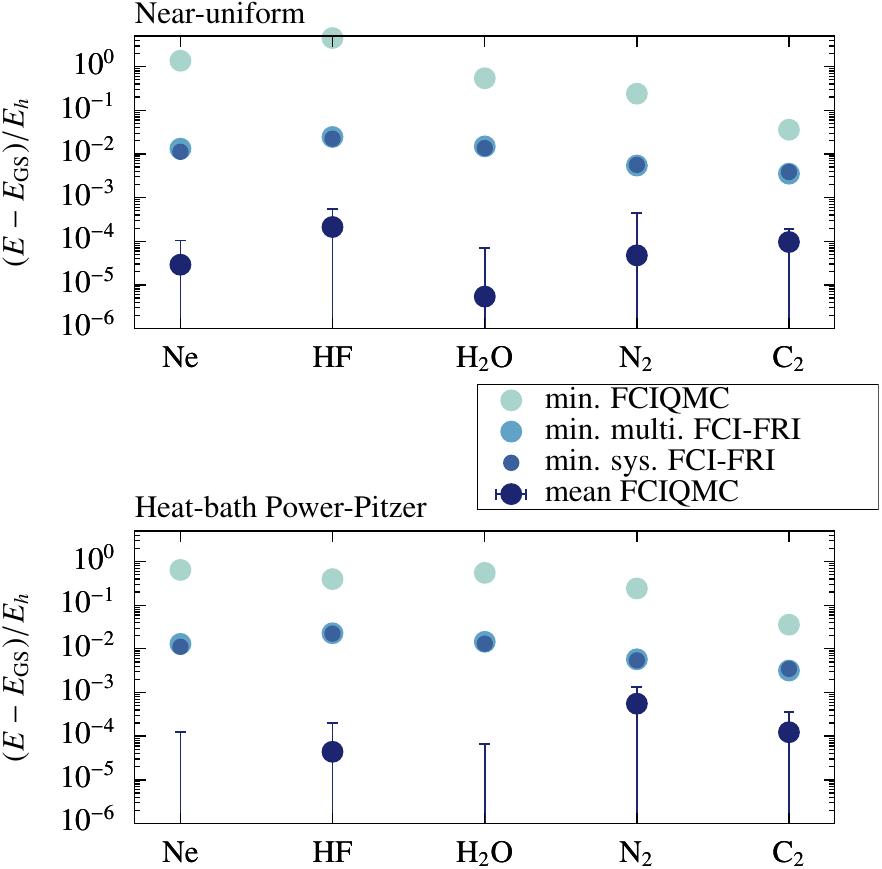}
\caption{The difference between the minimum variational energy estimate from each method and the exact FCI energy from Table \ref{tab:params}. Results from the FCIQMC, multinomial FCI-FRI, and systematic FCI-FRI methods, using both the near-uniform (top) and heat-bath Power-Pitzer (bottom) matrix factorization schemes, are shown for each of the molecular systems considered in this study. Mean energy differences from the FCIQMC method for each system are plotted for comparison. Error bars represent 95\% ($2\sigma_E$) confidence intervals.}
\label{fig:rayleigh}
\end{figure}

\subsection{Variational Energy Estimates}
Finally, we evaluate the possibility that the primary utility of the FCI-FRI methods
considered here is that they efficiently identify the most important Slater
determinant basis elements in the ground-state eigenvector. Variational Rayleigh
quotients (eq \ref{eq:rayEn}) for a subset of the iterates (i.e. every
100\textsuperscript{th} iterate) in each trajectory were calculated in addition
to the projected estimates used to obtain average  energies. If FCI-FRI is only an efficient search for significant basis elements, then we expect many of these Rayleigh quotients to be close to the ground-state energy.

We calculate the minimum Rayleigh quotient over both independent trajectories for each system considered. Differences between these minimum energies and the exact ground-state energies for each system are plotted in Figure~\ref{fig:rayleigh}. The mean energy difference from the original FCIQMC method is also plotted for comparison, with error bars denoting the corresponding 95\% confidence interval. For all methods and systems considered, this difference for the minimum Rayleigh quotient is more than an order of magnitude greater than the maximum of the FCIQMC confidence interval. The minimum Rayleigh quotients from FCIQMC are greater than those from the FCI-FRI methods considered and, for all systems except \ce{C2}, are also greater than the Hartree-Fock energy. 
This difference between the FCIQMC and FCI-FRI Rayleigh quotients can possibly be attributed to the lower-variance vector compression scheme employed in FCI-FRI. Even though the average of the FCIQMC iterates converges to the ground state to within a bias, the binomial integerization scheme used in FCIQMC displaces each iterate further from the ground state than in FCI-FRI.

These results indicate that none of the vectors from the FCIQMC or FCI-FRI trajectories are particularly close to the ground-state, as measured by the variational energy estimates. The facts that the average of each component of the solution vector converges quickly to its exact value, to within a controllable statistical bias, and that the projected estimator is linear in these components, rather than quadratic, are essential for the success of FCI-FRI methods. 

\section{Conclusions}
\label{sec:concl}
This paper describes several generic matrix and vector compression techniques within the FRI framework in the context of the FCI problem. Hierarchical approaches to matrix compression are discussed and shown to offer significant advantages over approaches that require enumerating all nonzero elements. Two examples of hierarchical factorization schemes for the FCI Hamiltonian matrix are presented, namely near-uniform and heat-bath Power-Pitzer. We describe how these various techniques can be combined in methods for calculating the FCI ground-state energy using power iteration, and we compare these ``FCI-FRI'' methods to FCIQMC in its original form.

Calculations on small molecules are used to compare the performance of these methods in terms of statistical efficiency, a metric inversely related to the square of the standard error. FCI-FRI calculations on the Ne atom demonstrate that using matrix compression in addition to vector compression can enable significant reductions in computational cost while only moderately decreasing the statistical efficiency.

We show that systematic matrix compression offers significant advantages over multinomial matrix compression, which has been used previously in FCIQMC. FCI-FRI calculations with systematic matrix compression applied to five small molecular systems are 11 to 45 times more efficient than those with multinomial compression, which are in turn 1.4 to 178 times more efficient than calculations performed using the original FCIQMC method.

The advantages of these stochastic methods over related deterministic compression methods are investigated. The error in a stochastic calculation on the Ne atom is nearly two orders of magnitude less than a deterministic calculation with comparable cost, which illustrates the importance of stochastically representing all components of the solution vector in the FCI Slater determinant space. Furthermore, by applying variational energy estimators to stochastic calculations performed on all molecular systems, we demonstrate the importance of averaging over many sparse, stochastic iterates in producing an accurate energy estimate. These features of stochastic methods and the results in this study suggest the applicability of FCI-FRI methods to strongly correlated systems with dense solution vectors.

Future research will investigate strategies for further improving the performance of FCI-FRI methods. We will develop implementations of these methods that exploit parallelism more effectively, possibly using techniques developed previously for FCIQMC. Due to the generality of the FRI framework, the compression techniques introduced here can be applied in tandem with the complementary initiator and semi-stochastic extensions to FCIQMC, which suggests an approach to further improving statistical efficiency. Additionally, examining the effect of the choice of parameters used in FCI-FRI calculations on the statistical efficiency may provide additional insight into how to optimize performance. For example, our results suggest that FCI-FRI methods allow more flexibility than FCIQMC in the choice of the parameter $\varepsilon$, which corresponds to the time step in imaginary time propagation. Varying $\varepsilon$ may affect the statistical efficiency of FCI-FRI methods. Furthermore, the number of nonzero elements in each matrix and vector compression in FCIQMC is determined by the number of walkers, whereas in FCI-FRI, these parameters can be varied independently. FCIQMC methods require a critical number of walkers to reliably converge to the ground-state energy. Our results suggest that using improved matrix compression schemes in FCI-FRI methods can reduce the number of matrix and vector elements required for convergence. Exploring these possibilities may facilitate the development of stochastic methods for quantum chemistry that are able to treat larger systems than currently possible.

\appendix
\section{Matrix Factorizations for Quantum Chemistry}
\label{sec:matFact}
This section describes two approaches to factoring the matrix $\mathbf{P}^{(\tau)}$, near-uniform and heat-bath Power-Pitzer (HB-PP). Elements in each matrix in the factorization are calculated using information from Hartree-Fock based on predetermined rules. The cost of evaluating these elements is greater for the HB-PP factorization than for near-uniform, although intermediate compression steps in the HB-PP scheme yield less statistical error than for near-uniform.

\subsection{Near-Uniform}
\label{sec:nearUniQ}
In the near-uniform factorization~\cite{Booth2014}, $\mathbf{P}^{(\tau)}$ is factored into the product $\mathbf{B} \mathbf{C}^{(\tau)} \mathbf{Q}$, where $\mathbf{Q}$ is factored further into a product of four matrices, $\mathbf{Q}^{(4)} \mathbf{Q}^{(3)} \mathbf{Q}^{(2)} \mathbf{Q}^{(1)}$. Elements of these four matrices can be calculated efficiently based on symmetry relationships between pairs of Slater determinants in the FCI basis.  Elements of $\mathbf{Q}$ differ from elements of $\mathbf{P}^{(\tau)}$, so multiplication by $\mathbf{C}^{(\tau)}$ compensates for this by multiplying by elements of $\mathbf{P}^{(\tau)}$ and dividing by elements of $\mathbf{Q}$. This ensures that elements of the product of matrix factors are equal to those of $\mathbf{P}^{(\tau)}$. Finally, multiplication by $\mathbf{B}$ sums elements corresponding to all excitations that contribute to each Slater determinant element of the final vector.

Each of the one-electron orbitals from a Hartree-Fock calculation can be assigned an associated irreducible representation (irrep) according to the symmetry of the system under consideration. This can encode spin symmetry (up or down), spatial (point group) symmetry, and, for crystalline systems, $k$-point symmetry. For each nonzero element in $\mathbf{H}$ corresponding to a single excitation from $\ket{{K}}$ to $\hat{c}^\dagger_a \hat{c}_i \ket{{K}}$, the irrep of orbital $i$ must equal that of orbital $a$, i.e. $\Gamma_i = \Gamma_a$. Because $\mathbf{P}^{(\tau)}$ is related to $\mathbf{H}$ by only a scalar factor and a shift by identity, its elements obey the same symmetry relationships. For double excitations, the direct product of irreps of the occupied orbitals, $\Gamma_i \otimes \Gamma_j$, must equal that of the virtual orbitals, $\Gamma_a \otimes \Gamma_b$, in order for the corresponding element of $\mathbf{H}$ to be nonzero. Excitations satisfying these symmetry constraints are termed symmetry-allowed excitations. %Elements in the near-uniform $\mathbf{Q}$ corresponding to symmetry-unallowed excitations are defined to be zero, as it is known a priori that their corresponding elements in $\mathbf{H}$ are zero.
Applying this factorization scheme requires an $O(N)$ operation per nonzero element in the current iterate to count the number of occupied and virtual orbitals with each irrep.

\begin{figure*}
\includegraphics[width=0.7\linewidth]{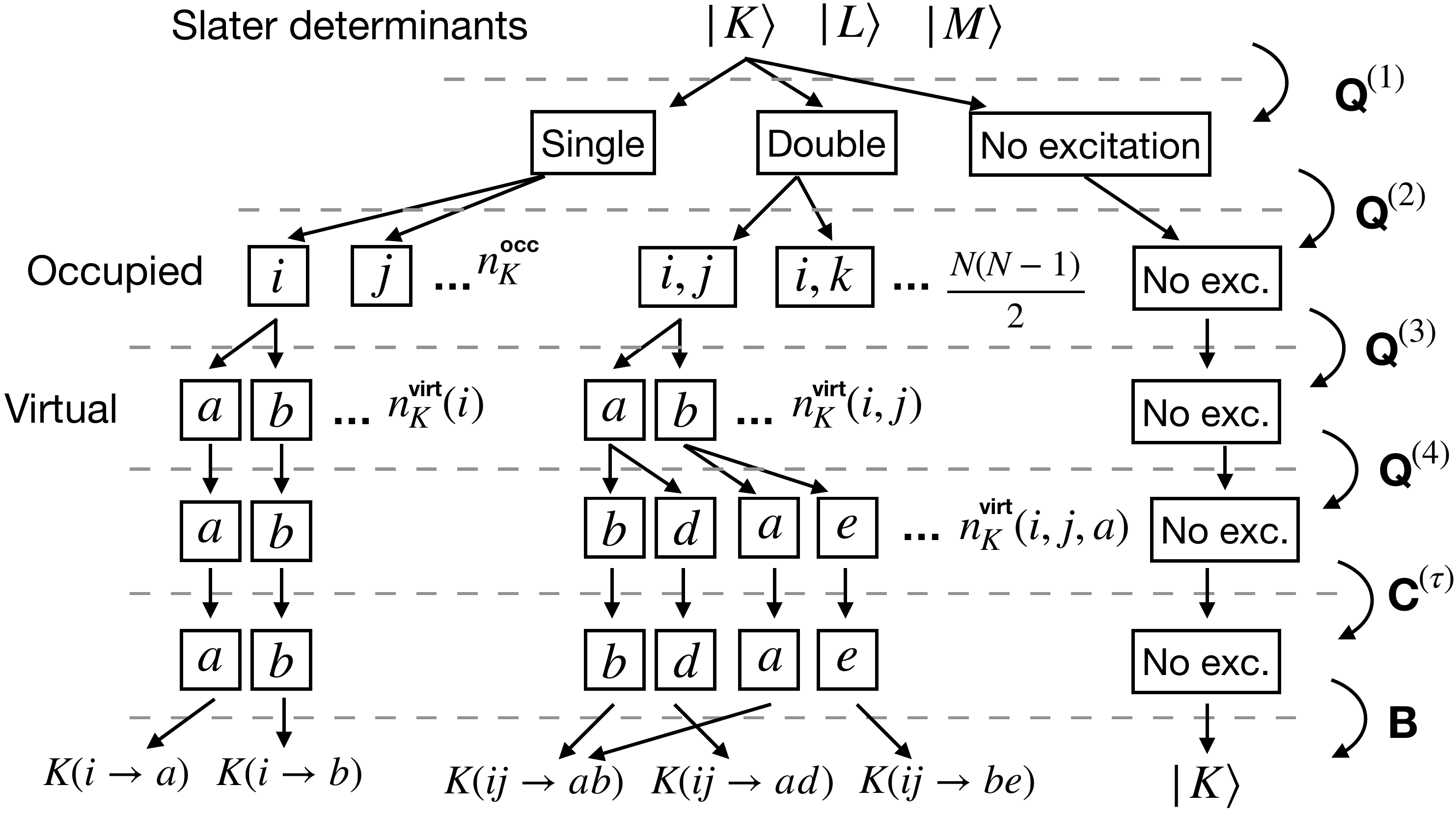}
\caption{The hierarchical structure of the near-uniform factorization of the matrix $\mathbf{P}^{(\tau)}$, showing the orbital indices used to index the row and column spaces of each matrix in the factorization.}
\label{fig:nearuni}
\end{figure*}

The matrices in this factorization map Slater determinant basis elements to excitations, indexed using multi-indices containing the orbitals involved in each excitation, and ultimately back to the determinants defined by these excitations. A schematic overview of these relationships is presented in Figure \ref{fig:nearuni}. The matrix $\mathbf{Q}^{(1)}$ has dimensions $(3 N_\text{FCI} \times N_\text{FCI})$, and its row space can be divided into three distinct subspaces. The first corresponds to single excitations, and each element is indexed using a multi-index $\lbrace K, 1 \rbrace$ denoting a generic single excitation from $\ket{K}$. Elements of $\mathbf{Q}^{(1)}$ in this subspace are given as
\begin{equation}
Q^{(1)}_{\lbrace K, 1 \rbrace, J} = \delta_{KJ} \frac{n_\text{s}}{n_\text{s} + n_\text{d}}
\end{equation}
where $\delta_{KJ}$ is a Kronecker delta, and $n_\text{s}$ and $n_\text{d}$ are the number of symmetry-allowed single and double excitations from a reference determinant in the FCI basis (typically Hartree-Fock). The second subspace contains generic double excitations and has elements given as
\begin{equation}
Q^{(1)}_{\lbrace K, 2 \rbrace, J} = \delta_{KJ} \frac{n_\text{d}}{n_\text{s} + n_\text{d}}
\end{equation}
Elements in the third subspace, indexed as $\lbrace K, 0 \rbrace$, will be mapped back to their original Slater determinant $\ket{K}$ by the final matrix multiplication in the factorization. These must be considered separately in intermediate steps, as will be explained in the discussion of compression below. These ``no excitation'' elements are given as
\begin{equation}
Q^{(1)}_{\lbrace K, 0 \rbrace, J} = \delta_{KJ}
\end{equation}

The subsequent matrices in the factorization map generic single and double excitations from the row space of $\mathbf{Q}^{(1)}$ to specific single and double excitations. This begins with multiplication by $\mathbf{Q}^{(2)}$, which maps to the specific occupied orbitals in these excitations. Single-excitation elements in this matrix are nonzero only for symmetry-allowed choices of occupied orbitals $i$. An occupied orbital in $\ket{K}$ is symmetry-allowed if there is at least one virtual orbital of the same symmetry in $\ket{K}$. The number of such orbitals in $\ket{K}$ is denoted $n_K^\text{occ}$. These single excitation elements in $\mathbf{Q}^{(2)}$ are
\begin{equation}
Q^{(2)}_{\lbrace K, 1, i \rbrace, \lbrace K, 1 \rbrace} = \left( n_K^\text{occ} \right)^{-1}
\end{equation}
Double excitation elements are nonzero for all of the $N(N-1)/2$ unique pairs of occupied orbitals $(i,j)$ in $\ket{K}$, regardless of whether they have corresponding symmetry-allowed pairs of virtual orbitals. These elements are
\begin{equation}
Q^{(2)}_{\lbrace K, 2, (i,j) \rbrace, \lbrace K, 2 \rbrace} = 2 N^{-1} (N - 1)^{-1}
\end{equation}
The orbitals $(i,j)$ are grouped in the multi-index to indicate that their order is irrelevant to the indexing. As above, ``no excitation'' elements of  $\mathbf{Q}^{(2)}$ are 1, i.e.
\begin{equation}
Q^{(2)}_{\lbrace K, 0 \rbrace, \lbrace K, 0 \rbrace} = 1
\end{equation}
All other elements of $\mathbf{Q}^{(2)}$ are 0.

Single-excitation elements in $\mathbf{Q}^{(3)}$ map a virtual orbital to each excitation. Elements for symmetry-allowed virtual orbitals $a$ are:
\begin{equation}
Q^{(3)}_{\lbrace K, 1, i, a \rbrace, \lbrace K, 1, i \rbrace} = \left( n_K^\text{virt}(i) \right)^{-1}
\end{equation}
where $n_K^\text{virt}(i)$ is the number of virtual orbitals in $\ket{K}$ with the same symmetry as $i$. Double excitation elements are defined for symmetry-allowed virtual orbitals $a$, i.e. those for which there exists at least one virtual orbital $b$ that satisfies $\Gamma_i \otimes \Gamma_j = \Gamma_a \otimes \Gamma_b$:
\begin{equation}
Q^{(3)}_{\lbrace K, 2, (i,j), a \rbrace, \lbrace K, 2, (i,j) \rbrace} = \left[ n_K^\text{virt}(i,j) \right]^{-1}
\end{equation}
where $n_K^\text{virt}(i,j)$ is the number of symmetry-allowed virtual orbitals given an occupied pair $(i,j)$. ``No excitation'' elements of $\mathbf{Q}^{(3)}$ are 1, and all other elements are 0.

Since single excitations are specified completely by the occupied and virtual orbitals in the row-space indices of $\mathbf{Q}^{(3)}$, single-excitation elements of $\mathbf{Q}^{(4)}$ map these excitations to themselves, as follows
\begin{equation}
\label{eq:singleId}
Q^{(4)}_{\lbrace K, 1, i, a \rbrace, \lbrace K, 1, i, a \rbrace} = 1
\end{equation}
Double-excitation elements for symmetry-allowed virtual orbitals $b$ are
\begin{equation}
Q^{(4)}_{\lbrace K, 2, (i,j), a, b \rbrace, \lbrace K, 2, (i,j), a \rbrace} = \left[n_K^\text{virt}(i,j,a) \right]^{-1}
\end{equation}
where $n_K^\text{virt}(i,j,a)$ denotes the number of symmetry-allowed virtual orbitals in $\ket{K}$ given $i$, $j$, and $a$. Note that all of these orbitals $b$ have the same irrep, since there is only one irrep $\Gamma_b$ in the system's point group that satisfies $\Gamma_i \otimes \Gamma_j = \Gamma_a \otimes \Gamma_b$.

The matrix $\mathbf{C}^{(\tau)}$, which ensures that the factorization is equal to $\mathbf{P}^{(\tau)}$, is a diagonal square matrix. Its single-excitation elements are
\begin{equation}
C^{(\tau)}_{\lbrace K,1,i,a \rbrace, \lbrace K,1,i,a \rbrace} = \frac{P^{(\tau)}_K(i \rightarrow a)}{Q^{(1)}_{\lbrace K, 1 \rbrace, K} Q^{(2)}_{\lbrace K, 1, i \rbrace, \lbrace K, 1 \rbrace} Q^{(3)}_{\lbrace K, 1, i, a \rbrace, \lbrace K, 1, i \rbrace}}
\end{equation}
The denominator corresponds to the ``generation probability'' in the original description of FCIQMC. Double excitation elements in $\mathbf{C}^{(\tau)}$ are
\begin{widetext}
\begin{multline}
C^{(\tau)}_{\lbrace K, 2, (i,j), a, b \rbrace, \lbrace K, 2, (i,j), a, b \rbrace} = P^{(\tau)}_K(ij \rightarrow ab) \left(Q^{(1)}_{\lbrace K, 2 \rbrace, K} Q^{(2)}_{\lbrace K, 2, (i,j) \rbrace, \lbrace K, 2 \rbrace} \right)^{-1} \left(Q^{(3)}_{\lbrace K, 2, (i,j), a \rbrace, \lbrace K, 2, (i,j) \rbrace} Q^{(4)}_{\lbrace K, 2, (i,j), a, b \rbrace, \lbrace K, 2, (i,j), a \rbrace} + \right. \\
\left. Q^{(3)}_{\lbrace K, 2, (i,j), b \rbrace, \lbrace K, 2, (i,j) \rbrace} Q^{(4)}_{\lbrace K, 2, (i,j), b, a \rbrace, \lbrace K, 2, (i,j), b \rbrace} \right)^{-1}
\end{multline}
\end{widetext}
The sum of terms in the denominator of this expression accounts for the fact that there are two elements in the row space of $\mathbf{C}^{(\tau)}$, i.e. $\lbrace K, 2, (i,j),a,b \rbrace$ and $\lbrace K, 2, (i,j),b,a \rbrace$, corresponding to each double excitation, i.e. $K(ij \rightarrow ab)$. These will be summed after multiplication by the final matrix in the factorization, $\mathbf{B}$. Elements in $\mathbf{C}^{(\tau)}$ corresponding to ``no excitation'' elements in the basis are given as their corresponding diagonal elements in $\mathbf{P}^{(\tau)}$:
\begin{equation}
C^{(\tau)}_{\lbrace K, 0 \rbrace, \lbrace K, 0 \rbrace}  = P^{(\tau)}_{KK}
\end{equation}

Multiplication by $\mathbf{B}$ sums contributions from the row space of $\mathbf{C}^{(\tau)}$ that map to the same Slater determinant basis element. Because there are many elements in this space that map to the same determinant, the row dimension of $\mathbf{B}$ is smaller than the column dimension. Elements for double excitations are
\begin{equation}
B_{K(ij \rightarrow  ab), \lbrace K, 2, (i,j),a,b \rbrace} = B_{ K(ij \rightarrow  ab) , \lbrace K, 2, (i,j),b,a \rbrace} = 1
\end{equation}
and those for single excitations are
\begin{equation}
B_{K(i \rightarrow a), \lbrace K, 1, i, a \rbrace} = 1
\end{equation}
``No excitation'' elements are mapped back to the determinant from which they originated, i.e.
\begin{equation}
B_{K, \lbrace K, 0 \rbrace} = 1
\end{equation}
This mapping can be performed efficiently using a hashing algorithm~\cite{Booth2014}, at $O(N_\text{mat})$ cost, where $N_\text{mat}$ is the number of elements selected from the matrix. In our current implementations of FCIQMC and FCI-FRI methods, a simpler $O(N_\text{mat} \log m)$ binary search is used instead, where $m$ is the number of nonzero iterate elements.

\begin{figure*}
\includegraphics[width=0.7\linewidth]{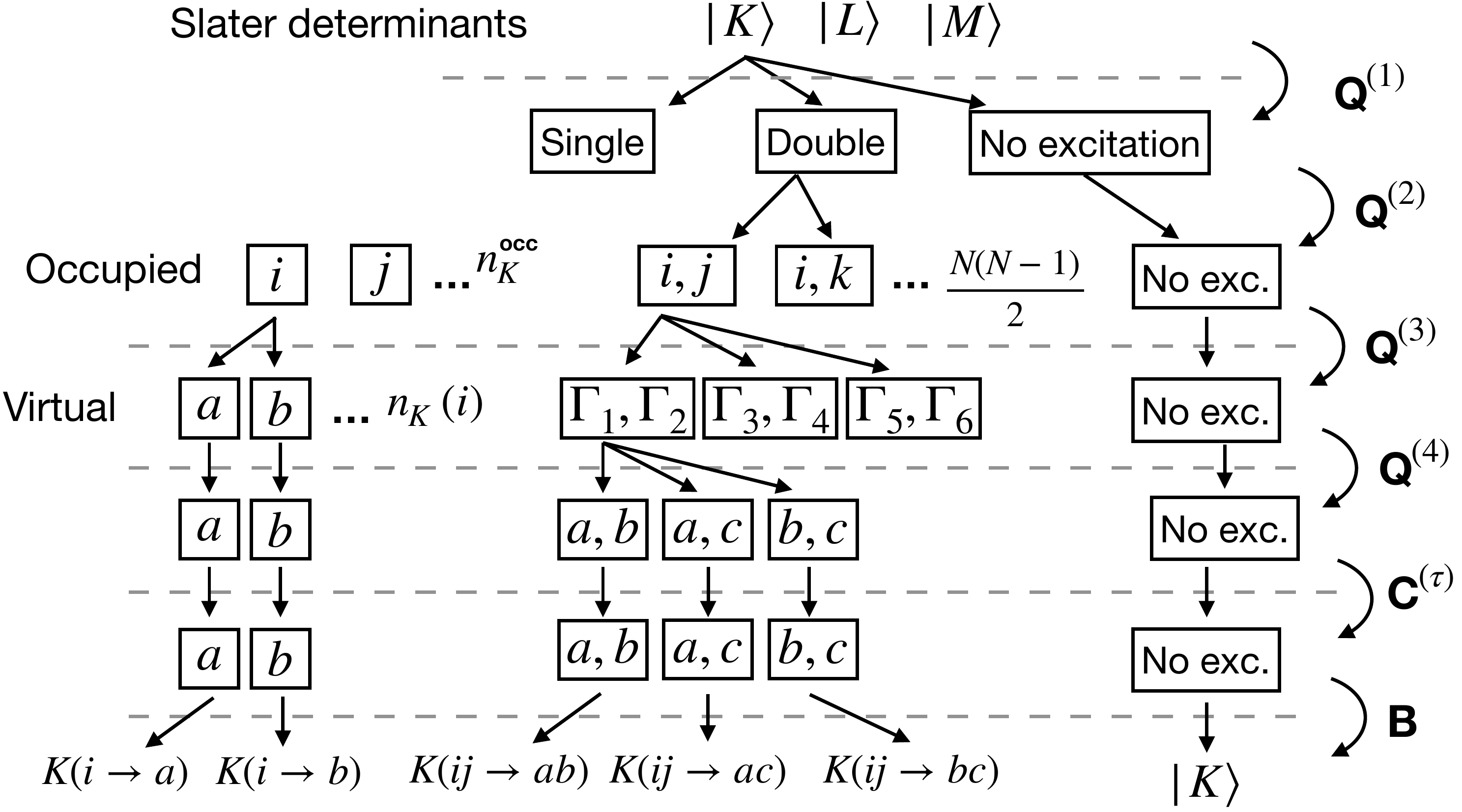}
\caption{An alternative near-uniform factorization structure for the matrix $\mathbf{P}^{(\tau)}$. Unlike in Figure \ref{fig:nearuni}, double excitations have unique indices, making this factorization more amenable to systematic compression.}
\label{fig:nearunisys}
\end{figure*}

The selection of excitations from the near-uniform distribution in FCIQMC can be understood as a particular multinomial compression technique applied to the factorization scheme discussed above. Systematic compression could be applied analogously. However, one of the primary advantages of systematic compression is that it can be performed such that the elements selected from the vector are unique, thereby yielding less statistical error. This benefit is somewhat diminished when using the factorization described above, since the indices $\lbrace K, 2, (i,j),a,b \rbrace$ and $\lbrace K, 2, (i,j),b,a \rbrace$ indicate the same double excitation but are treated separately until multiplication by $\mathbf{B}$. We therefore designed an alternative factorization scheme to address this (Figure \ref{fig:nearunisys}). Elements of the matrix $\mathbf{B} \mathbf{Q}$ in this alternative scheme are equal to those in the original scheme; the difference only arises in how double excitation elements are defined and indexed in $\mathbf{Q}^{(3)}$ and $\mathbf{Q}^{(4)}$. Pairs of virtual orbitals and pairs of symmetry elements in the system's point group are used instead of individual virtual orbitals to index these elements. Applying the FCIQMC compression technique with the alternative scheme would yield the same statistical error; its advantages are realized only when using systematic compression.

Elements of the matrix $\mathbf{Q}^{(3)}$ in this alternative factorization were obtained by summing double excitation elements in the matrix product $\mathbf{Q}^{(3)} \mathbf{Q}^{(4)}$ from the above factorization corresponding to pairs of irreps $(\Gamma_x, \Gamma_y)$. If the irreps of the occupied orbitals in a double excitation are equal $(\Gamma_i = \Gamma_j)$, then the irreps of the virtual orbitals must also be equal to satisfy the symmetry conditions described above. Double excitation elements of $\mathbf{Q}^{(3)}$ corresponding to such pairs of occupied orbitals are
\begin{multline}
Q^{(3)}_{\lbrace K, 2, (i,j),(x,x) \rbrace, \lbrace K, 2, (i,j) \rbrace} =  \\
 \begin{cases}
n_K^\text{virt} (\Gamma_x) n_K^\text{virt}(i, j)^{-1} & n_K^\text{virt}(\Gamma_x) > 1, \Gamma_i \otimes \Gamma_j = \Gamma_x \otimes \Gamma_x \\
0 & \text{otherwise}
\end{cases}
\end{multline}
Here $x$ denotes a symmetry element in the system's point group, $\Gamma_x$ is its associated irrep, and $n_K^\text{virt} (\Gamma_x)$ denotes the number of virtual orbitals in $\ket{K}$ with irrep $\Gamma_x$. If $\Gamma_i \neq \Gamma_j$, the corresponding elements of $\mathbf{Q}^{(3)}$ are
\begin{multline}
Q^{(3)}_{\lbrace K, 2, (i, j), (x, y) \rbrace, \lbrace K, 2, (i,j) \rbrace} = \\ 
\begin{cases}
\frac{n_K^\text{virt}(\Gamma_x) + n_K^\text{virt}(\Gamma_y)} {n_K^\text{virt}(i, j)} & n_K^\text{virt}(\Gamma_x) > 0, n_K^\text{virt}(\Gamma_y) > 0, \Gamma_i \otimes \Gamma_j = \Gamma_x \otimes \Gamma_y \\
0 & \text{otherwise}
\end{cases}
\end{multline}

Double excitation elements in $\mathbf{Q}^{(4)}$ are given as the reciprocal of the number of virtual orbital pairs within each irrep pair. For pairs of virtual orbitals with the same irrep,
\begin{multline}
Q^{(4)}_{\lbrace K, 2, (i, j), (a, b) \rbrace, \lbrace K, 2, (i,j),(x,x) \rbrace} =  \\
\begin{cases}
\frac{2}{n_K^\text{virt}(\Gamma_x) \left[n_K^\text{virt}(\Gamma_x) - 1 \right]} & n_K^\text{virt}(\Gamma_x) > 1, \Gamma_a = \Gamma_b = \Gamma_x \\
0 & \text{otherwise}
\end{cases}
\end{multline}
If instead $\Gamma_a \neq \Gamma_b$, the elements are 
\begin{multline}
Q^{(4)}_{\lbrace K, 2, (i, j), (a, b) \rbrace, \lbrace K, 2, (i, j), (x, y) \rbrace} =  \\ \begin{cases}
n_K^\text{virt}(\Gamma_x)^{-1} n_K^\text{virt}(\Gamma_y)^{-1} & n_K^\text{virt}(\Gamma_x) > 0, n_K^\text{virt}(\Gamma_y) > 0, \Gamma_a = \Gamma_x, \Gamma_b = \Gamma_y \\
0 & \text{otherwise}
\end{cases}
\end{multline}
Except for the different indexing scheme for virtual orbitals in double excitations, elements in $\mathbf{C}^{(\tau)}$ and $\mathbf{B}$ are defined as above. Consequently, the elements of $\mathbf{C}^{(\tau)}$ are as uniform in magnitude as in the factorization scheme above. Compression of either near-uniform factorization scheme can be performed at approximately $O(N_\text{mat})$ cost.

\subsection{Heat-Bath Power-Pitzer}
In the above factorization, symmetry information is used to facilitate the efficient calculation of elements in the first four matrices, and discrepancies between products of these elements and elements of $\mathbf{P}^{(\tau)}$ are eliminated through multiplication by $\mathbf{C}^{(\tau)}$. Less error is introduced by stochastic compression of this factorization when $\mathbf{Q}$ is closer to $\mathbf{P}^{(\tau)}$, i.e. when the magnitudes of elements of $\mathbf{C}^{(\tau)}$ are more uniform~\cite{Holmes2016, Neufeld2019}. The heat-bath Power-Pitzer (HB-PP) factorization is designed to achieve more uniformity in these elements by using information from the Hamiltonian matrix in constructing $\mathbf{Q}$, which is factored into a product of five matrices, $\mathbf{Q}^{(5)} \mathbf{Q}^{(4)} \mathbf{Q}^{(3)} \mathbf{Q}^{(2)} \mathbf{Q}^{(1)}$. Elements in these matrices are indexed by individual orbitals rather than unique pairs of orbitals or symmetry elements. Because it is expensive to incorporate information about single-excitation Hamiltonian elements into these matrices, due to the $O(N)$ cost of evaluating each element, single-excitation elements in the factors of $\mathbf{Q}$ are defined exactly as in the near-uniform case. The same is true for ``no excitation'' elements, for reasons that will be made apparent in Appendix \ref{sec:FCIcomp}.

Elements corresponding to double excitations in $\mathbf{Q}^{(1)}$ are also defined as in the near-uniform case. Double-excitation elements in subsequent matrices are defined in terms of a matrix $\mathbf{D}$ and vector $\mathbf{S}$. Elements of $\mathbf{D}$ approximate the sum of magnitudes of double excitation elements in the Hamiltonian corresponding to a particular pair of occupied orbitals,\cite{Holmes2016}
\begin{equation}
D_{pq} = \begin{cases}
\sum_{r,s \notin \lbrace p, q \rbrace} \left\lvert \mel{p q }{}{ r s} \right\rvert & p \neq q \\
0 & p=q
\end{cases}
\end{equation}
where $\mel{p q }{}{ r s}$ is an antisymmetrized two-electron integral obtained from the Hartree-Fock calculation. The exact sum for each determinant depends on which
orbitals are occupied, so it is approximated by an unrestricted sum over all other orbitals in the Hartree-Fock basis. Analogously, elements of $\mathbf{S}$ approximate this sum for a single occupied orbital,
\begin{equation}
S_p = \sum_{q} D_{pq}
\end{equation}
The primary advantage of defining $\mathbf{S}$ and $\mathbf{D}$ by unrestricted sums is that they can be computed and stored in memory at the beginning of the simulation, at a memory cost of $O(M^2)$ and a CPU cost of $O(M^4)$.

The row spaces of $\mathbf{Q}^{(2)}$ and $\mathbf{Q}^{(3)}$ are indexed by multi-indices containing individual occupied orbitals instead of unique pairs of occupied orbitals. Elements in $\mathbf{Q}^{(2)}$ are
\begin{equation}
Q^{(2)}_{\lbrace K, 2, i \rbrace, \lbrace K, 2 \rbrace} = \frac{S_i}{\sum_{j \in \text{occ}} S_j}
\end{equation}
and those in $\mathbf{Q}^{(3)}$ are
\begin{equation}
Q^{(3)}_{\lbrace K, 2, i, j \rbrace, \lbrace K, 2, i \rbrace} = \frac{D_{ij}}{\sum_{j' \in \text{occ}} D_{ij'}}
\end{equation}
As a consequence of this indexing scheme, pairs of elements in $\mathbf{Q}^{(3)}$ in which the order of the occupied orbitals is reversed are not necessarily equal, i.e.
\begin{equation}
Q^{(3)}_{\lbrace K, 2, i, j \rbrace, \lbrace K, 2, i \rbrace} \neq Q^{(3)}_{\lbrace K, 2, j, i \rbrace, \lbrace K, 2, i \rbrace}
\end{equation}

Elements in the next matrix corresponding to double excitations with one virtual orbital are defined as
\begin{equation}
Q^{(4)}_{\lbrace K, 2, i, j, a \rbrace, \lbrace K, 2, i, j \rbrace} = \frac{|\braket{i a }{a i}|^{1/2}}{\sum_{c \in \lbrace \text{virt} \rbrace} |\braket{i c }{ c i}|^{1/2}}
\end{equation}
where $\braket{i a }{ a i}$ represents a two-electron exchange integral. Note that if the spins of orbitals $i$ and $a$ differ, this integral is 0. The sum in the denominator includes all virtual orbitals in $\ket{K}$. Elements in $\mathbf{Q}^{(5)}$ are indexed by a second virtual orbital and are defined as
\begin{equation}
Q^{(5)}_{\lbrace K, 2, i, j, a, b \rbrace, \lbrace K, 2, i, j, a \rbrace} = \frac{|\braket{j b }{ b j}|^{1/2} \delta_{\Gamma_b \otimes \Gamma_a, \Gamma_i \otimes \Gamma_j}}{\sum_c | \braket{j c }{ c j}|^{1/2} \delta_{\Gamma_c \otimes \Gamma_a, \Gamma_i \otimes \Gamma_j}}
\end{equation}
where the Kronecker deltas enforce the symmetry condition for double excitations described in Section \ref{sec:nearUniQ}, and the sum includes \textit{all} orbitals in the basis, including those occupied in $\ket{K}$. Elements in $\mathbf{Q}^{(5)}$ corresponding to single excitations are defined in analogy to eq \ref{eq:singleId}.

Elements of the matrix $\mathbf{C}^{(\tau)}$ are
\begin{widetext}
\begin{multline}
C^{(\tau)}_{\lbrace K, i, j, a, b\rbrace, \lbrace K, 2, i, j, a, b \rbrace} = P^{(\tau)}_K(ij \to ab) \left(Q^{(1)}_{\lbrace K, 2 \rbrace, K} \right)^{-1} 
\left[ Q^{(2)}_{\lbrace K, 2, i \rbrace, \lbrace K, 2 \rbrace} Q^{(3)}_{\lbrace K, 2, i, j \rbrace, \lbrace K, 2, i \rbrace} \left( Q^{(4)}_{\lbrace K, 2, i, j, a \rbrace, \lbrace K, 2, i,j \rbrace} Q^{(5)}_{\lbrace K, 2, i, j, a,b \rbrace, \lbrace K, 2, i,j,b \rbrace} + \right. \right.\\
\left. Q^{(4)}_{\lbrace K, 2, i, j, b \rbrace, \lbrace K, 2, i,j \rbrace} Q^{(5)}_{\lbrace K, 2, i, j, b,a \rbrace, \lbrace K, 2, i,j,a \rbrace} \right) + \\
Q^{(2)}_{\lbrace K, 2, j \rbrace, \lbrace K, 2 \rbrace} Q^{(3)}_{\lbrace K, 2, j,i \rbrace, \lbrace K, 2, j \rbrace} \left( Q^{(4)}_{\lbrace K, 2, j,i, a \rbrace, \lbrace K, 2, j,i \rbrace} Q^{(5)}_{\lbrace K, 2, j,i, a,b \rbrace, \lbrace K, 2, j,i,b \rbrace} + \right.\\
\left. \left. Q^{(4)}_{\lbrace K, 2, j,i, b \rbrace, \lbrace K, 2, j,i \rbrace} Q^{(5)}_{\lbrace K, 2, j,i, b,a \rbrace, \lbrace K, 2, j,i,a \rbrace} \right) \right]^{-1} 
\end{multline}
\end{widetext}
where the four terms in the sum account for the four different orders in which the orbitals for the double excitation can be chosen. The matrix $\mathbf{B}$ is defined analogously to the near-uniform factorization. The cost of performing the compressions for the HB-PP scheme scales as $O(M N_\text{mat})$.

\section{Compression Schemes in FCIQMC and FCI-FRI}
\label{sec:FCIcomp}
In principle, any compression scheme could be used to compress the intermediate vectors generated after each matrix multiplication in the hierarchical factorization schemes described above. This section describes the specific schemes used in the original FCIQMC method, as well as in multinomial and systematic FCI-FRI. Previously, FCIQMC has been described in terms of a sequence of ``spawning,'' ``death/cloning,'' and ``annihilation'' steps. This section presents an alternative interpretation of the method using the language of FRI.

Different subspaces of the intermediate vectors are treated differently in the compression schemes used in each of these methods. In each vector obtained after multiplying by the factors of $\mathbf{Q}$, ``no excitation'' elements are preserved exactly in all methods considered in this study. This is because diagonal elements of $\mathbf{P}^{(\tau)}$ are often significantly greater in magnitude than off-diagonal elements, provided that $\varepsilon$ is sufficiently small (eq \ref{eq:Puncomp}). 

In FCIQMC, the remaining portions of the vectors are compressed using multinomial sampling, without exact preservation of elements, with the added constraint that certain numbers of samples are allocated to each subspace. The number of samples allocated to an arbitrary subspace $w$ is denoted $n_w$. The number of samples allocated to the space of excitations associated with each Slater determinant is given as
\begin{equation}
n_{\lbrace K \rbrace} = |v_K|
\end{equation}
The number of elements in each single- and double-excitation subspace is determined by counting the number of samples in each subspace during multinomial compression of the vector $\mathbf{Q}^{(1)} \mathbf{v^{(\tau)}}$. The numbers in the remaining subspaces are calculated analogously following each matrix multiplication. The total number of samples used in each compression is denoted $N_\text{mat}$. Approaches to performing this sampling efficiently for the near-uniform and HB-PP factorizations are described in refs \citenum{Booth2014}, \citenum{Holmes2016}, and \citenum{Neufeld2019}.

Compression in FCIQMC is performed differently following multiplication by $\mathbf{C}^{(\tau)}$ in both factorizations, so that each element in the resulting vector is an integer. If $\mathbf{x}'$ denotes the vector obtained after multiplication by this matrix, ``no excitation'' elements in the compressed vector are given as
\begin{equation}
\label{eq:clone}
\Phi (\mathbf{x}')_{\lbrace K, 0 \rbrace} = \sum_{i=1}^{|v_K|} \text{bin}^{(i)}\left( x'_{\lbrace K, 0 \rbrace} v_K^{-1} \right)
\end{equation}
The function $\text{bin}^{(i)}(x)$ denotes the binomial integerization of a number $x$, defined as
\begin{equation}
\text{bin}^{(i)}(x) = \lfloor x + r^{(i)} \rfloor
\end{equation}
where $r^{(i)}$ is a random number chosen uniformly on the interval $(0, 1)$. This function preserves its argument in expectation, i.e. $\text{E}[\text{bin}^{(i)}(x)] = x$. Different values of the superscript $i$ correspond to independent random numbers. The argument of this function in eq \ref{eq:clone}  is related to the ``death/cloning probability'' in previous presentations of FCIQMC, and performing the sampling corresponds to the ``diagonal death/cloning'' step. Other elements in the compressed vector, corresponding to off-diagonal elements in $\mathbf{P}^{(\tau)}$, are
\begin{equation}
\Phi (\mathbf{x}')_i = \sum_{j=1}^{n_i} \text{bin} \left( x'_i n_i^{-1} \right)
\end{equation}
where the index $i$ indicates an excitation, e.g. $\lbrace K, (i, j),a,b \rbrace$ or $\lbrace K, i, a \rbrace$. Here the argument of the binomial integerization function corresponds to the ``spawning probability'' in FCIQMC. The resulting vector is sparse because many elements are set to zero by the binomial integerization function. The number of nonzero elements is random, unlike the systematic vector compression. Multiplication by $\mathbf{B}$ constitutes the ``annihilation'' step in FCIQMC, as it involves summing elements that are mapped to the same Slater determinant basis element.

In multinomial FCI-FRI, multinomial compression is also used to compress the first few intermediate vectors. Because the elements of $\mathbf{v}^{(\tau)}$ are not necessarily integers, a separate systematic sampling procedure is applied to determine the number of elements $n_{\lbrace K \rbrace}$ to sample from the subspace associated with each Slater determinant. The magnitudes of elements in $\mathbf{v}^{(\tau)}$ are normalized to obtain the probabilities $\lbrace p_i \rbrace$ used in systematic sampling, and a constraint is added: for all $K$ for which $|v^{(\tau)}_K| > 0$, $n_{\lbrace K \rbrace} > 0$. In contrast to FCIQMC, compression is not performed after multiplication by $\mathbf{C}^{(\tau)}$, so the elements summed during multiplication by $\mathbf{B}$ are not necessarily integers. In systematic FCI-FRI, the first few intermediate vectors are compressed systematically to $N_\text{mat}$ elements instead of multinomially, preserving $\rho$ elements exactly in each compression according to eq \ref{eq:rhoCriterion}. Unlike in multinomial compression, the constraint that a certain number of elements are selected from each subspace is not imposed. Because the order of elements determines which elements are chosen in systematic compression, elements are ordered consistently in each iteration, first by the Slater determinant index, then by the type of excitation (single, double, or ``no excitation''), then by the occupied and virtual orbital(s).

\begin{acknowledgments}
We thank Aaron Dinner and Sandeep Sharma for useful discussions about this work, and we thank Anthony Scemama for noting an error in the geometry of H$_2$O used in calculations in a preprint.
S.M.G.~and T.C.B.~were supported by start-up funding from the University
of Chicago and by the Flatiron Institute. The Flatiron Institute is a division of the Simons Foundation. R.J.W.~and J.W.~were supported by the
Advanced Scientific Computing Research program through award DE-SC0014205.
R.J.W.~was also supported by NSF RTG award 1547396 at the University of Chicago
and by a MacCracken Fellowship and NSF RTG award 1646339 at New York University. 
Calculations were performed with resources provided by the University of
Chicago Research Computing Center.
\end{acknowledgments}

\section*{References}
\bibliography{Sams_refs,software}

\end{document}